\documentclass[manuscript]{aastex}
\usepackage{times}
\usepackage{epsfig}
\usepackage{natbib}
\usepackage{amssymb}
\usepackage{amsmath}
\usepackage{graphicx}
\usepackage{graphics}
\usepackage{rotating}
\usepackage{fancyhdr}
\usepackage{lscape}
\usepackage{calc}
\usepackage{float}                      
\usepackage{dcolumn}                    
\usepackage{multirow}
\usepackage{longtable}
\usepackage{lscape}
\usepackage{subfigure}
\usepackage{appendix}
\usepackage{epstopdf}
\usepackage{array}

\def\be{\begin{equation}}\def\bea{\begin{eqnarray}}\def\beaa{\begin{eqnarray*}}
  \def\ee{\end{equation}}  \def\eea{\end{eqnarray}}  \def\eeaa{\end{eqnarray*}}

\def\fun#1#2{\lower3.6pt\vbox{\baselineskip0pt\lineskip.9pt
        \ialign{$\mathsurround=0pt#1\hfill##\hfil$\crcr#2\crcr\sim\crcr}}}

\newcommand{\aspcs}{Astronomical Society of the Pacific Conference Series}
\newcommand{\spie}{Society of Photo-Optical Instrumentation Engineers Conference Series}
\newcommand{\atel}{The Astronomer's Telegram}
\newcommand{\iaus}{IAU Symposium}
\newcommand{\aipc}{American Institute of Physics Conference Series}
\newcommand{\BaltA}{Baltic Astronomy}
\newcommand{\basi}{Bulletin of the Astronomical Society of India}

\slugcomment{Submitted to AJ}

\shorttitle{T Pyx (2011) from 0.8-250 Days after Discovery}
\shortauthors{Surina et al.}

\begin{document}


\title{The Detailed Photometric
and Spectroscopic Study of the 2011 Outburst \\of the Recurrent Nova T Pyxidis \\
    from 0.8 to 250 Days after Discovery}


\author{F. Surina}
\affil{Astrophysics Research Institute, Liverpool John Moores University\\ Liverpool Science Park, IC2 Building, 146 Brownlow Hill, Liverpool, L3 5RF, UK}
\email{mfs@astro.livjm.ac.uk}
\author{R. A. Hounsell}
\affil{Space Telescope Science Institute, 3700 San Martin Drive, Baltimore, MD 21218 410-338-4974}
\author{M. F. Bode, M. J. Darnley, D. J. Harman}
\affil{Astrophysics Research Institute, Liverpool John Moores University\\ Liverpool Science Park, IC2 Building, 146 Brownlow Hill, Liverpool, L3 5RF, UK}
\and
\author{F. M. Walter}
\affil{Department of Physics and Astronomy, Stony Brook University, Stony Brook, NY 11794-3800}

\begin{abstract}
We investigated the optical light curve of T Pyx during its 2011 outburst through compiling a database of Solar Mass Ejection Imager (SMEI) and AAVSO observations. The SMEI light curve, providing unprecedented detail covering $t$=1.5-49 days post-discovery, was divided into four phases based on the idealised nova optical light curve; the initial rise (1.5-3.3 days), the pre-maximum halt (3.3-13.3 days), the final rise (14.7-27.9 days), and the early decline (27.9 days - -). The SMEI light curve contains a strongly detected period of 1.44$\pm$0.05 days during the pre-maximum halt phase. These oscillations resemble those found in recent TNR models arising from instabilities in the expanding envelope. No spectral variations that mirror the light curve periodicity were found however. The marked dip at $t$$\sim$22-24 days just before light curve maximum at $t$=27.9 days may represent the same (shorter duration) phenomenon seen in other novae observed by SMEI and present in some model light curves. The spectra from the 2m Liverpool Telescope and SMARTS 1.5m telescope were obtained from $t$=0.8-80.7 and 155.1-249.9 days, covering the major phases of development. The nova was observed very early in its rise where a distinct high velocity ejection phase was evident with derived $V_{ej}$$\sim$4000 km s$^{-1}$ initially. A marked drop at $t$=5.7 days, and then a gradual increase occurred in derived $V_{ej}$ to stabilise at $\sim$1500 km s$^{-1}$ at the pre-maximum halt. Here we propose two different stages of mass loss, a short-lived phase occurring immediately after outburst and lasting $\sim$6 days followed by a more steadily evolving and higher mass loss phase. The overall spectral development follows that typical of a Classical Nova and comparison with the photometric behaviour reveals consistencies with the simple evolving pseudo-photosphere model of the nova outburst. Comparing optical spectra to X-ray and radio light curves, weak [Fe X] 6375\AA$ $ emission was marginally detected before the X-ray rise and was clearly present during the brightest phase of X-ray emission. If the onset of the X-ray phase and the start of the final decline in the optical are related to the cessation of significant mass loss, then this occurred at $t$$\sim$90-110 days.
\end{abstract}





\clearpage
\section{Introduction}
Classical novae (CNe) are interacting binary systems whose outbursts are powered by a thermonuclear runaway in accreted material on the surface of a white dwarf (WD). The secondary stars in such systems fill their Roche lobe and material is transferred onto the WD primary star via an accretion disc \citep{sta08}. Recurrent novae (RNe) are, by definition, CNe with multiple recorded outbursts and most contain evolved secondaries \citep{dar12}. There are 10 known Galactic RNe: T Pyx, IM Nor, CI Aql, V2487 Oph, U Sco, V394 CrA, T CrB, RS Oph, V745 Sco, and V3890 Sgr, given here in order of increasing orbital period (\citep{sch10b}, although the period of V2487 Oph is only approximately known). T Pyx ($\alpha$=9$^{h}$4$^{m}$41$^{s}\!.$50, $\delta$=$-$32$^{\circ}$22$'$47$''\!.$5) has the shortest known orbital period ($\sim$0$^{d}\!.$076 or 1$^{h}\!.$824) of any recurrent nova (\citealp[see][for a review]{anu08}) and the only RN that is below the cataclysmic variable (CV) period gap \citep{sch92,pat98,uth10}.



T Pyx has had previous observed outbursts in 1890, 1902, 1920, 1944, 1966/1967, and now 2011, with an extensive shell of ejected gas associated with these. The photometric and spectroscopic characteristics of the first outbursts were briefly presented in \citet{pay57} while the 1966 event was discussed post-outburst by \citet{cat69} and \citet{lan70}. All the outbursts have exhibited brightness fluctuations near the optical peak and the subsequent declines have been slow. The ejection velocity observed previously was $\sim$2000 km s$^{-1}$ in 1967 \citep{cat69,wil82} with $t_{3}$=63 days \citep{sch10b}. The distance has been revised most recently by \citet{sho11} to be $\ge$4.5 kpc, with a strict lower limit of 3.5 kpc which was the previously accepted value. Taking M$_{bol}$=$-$7.0 given by \citet{sch10b} gives $L_{bol}$ $\sim$ 2$\times$10$^{38}$ erg s$^{-1}$ at maximum light ($\sim$$L_{edd}$ for a 1M$_{\odot}$ WD).

T Pyx is unique for having an outburst light curve plateau in the decline stage which is different from the five other RNe (IM Nor, CI Aql, V2487 Oph, U Sco, and RS Oph) with observed (so-called `true') plateaus in their light curves. While the latter true plateaus are believed to be the result of the combination of the irradiation of the disc (by the supersoft emission from nuclear burning near the WD after the wind associated with the ejecta has stopped) and the steady light decline from the shell which leads to a flattening of the light curve until the nuclear burning turns off, the plateau in T Pyx is thought to arise from a different mechanism including the emission lines increasing in brightness \citep{hac00}. Unlike true plateaus that have the flat portion starting during an apparently final decline, T Pyx's plateau starts immediately after a sharp drop ($\sim$2.0 mag in 20 days after peak) and starts again in the final decline ($\sim$105 days after peak). This feature, of a plateau starting after a sharp drop, is also present in RS Oph; however, the presence of a Super Soft X-ray Source (SSS) and the lack of colour changes during the plateau phase in RS Oph indicates a true plateau \citep{hac08}. On the other hand, the shape of T Pyx's plateau in $B$ and $V$ band light curves is significantly different \citep{sch10b}.



In classical novae, the mass of the WD ($M_{WD}$) is typically around 1M$_{\odot}$ \citep{uth10}. For a RN, in order for the thermonuclear runaway (TNR) to occur on a short time scale, the WD has to be more massive (i.e. $M_{WD}$$\gtrsim$1M$_{\odot}$) and luminous with high $\dot{M}$ \citep{sta08a} implying that the secondary star has to be evolved. As the result of the unusually high $\dot{M}$, T Pyx is more luminous than CNe at quiescence in which the Roche lobe filling main-sequence donor star is transferring mass onto a high-mass WD. Unfortunately, $M_{WD}$ is not accurately known for this system.

Based on optical spectroscopy obtained at the Very Large Telescope and the Magellan Telescope, \citet{uth10} estimated a mass ratio $q$=0.2$\pm$0.03. Thus with the mass of the secondary star $M_{2}$=0.14$\pm$0.03M$_{\odot}$ derived from the main-sequence mass-radius relation of a 5-Gyr isochrone given by \citet{bar98} then the mass of the primary WD is $M_{WD}$=0.7$\pm$0.2M$_{\odot}$. This estimate of $M_{WD}$ is lower than that expected from previous theoretical studies which range from 1.30-1.37M$_{\odot}$ \citep{kat90}, 1.25-1.30M$_{\odot}$ \citep{sch10c}, and 1.25-1.4M$_{\odot}$ \citep{sel10}. Therefore \citet{uth10} turned the problem around by mentioning that if $M_{WD}$$>$1$M_{\odot}$ then $M_{2}$$>$0.2M$_{\odot}$ for their estimated $q$=0.2 mass ratio.


The orbital period at quiescence, $P$=0$^{d}.$076, is well established \citep{pat98,sch92,uth10} with an increasing period $\dot{P}=6\times10^{-10}$ \citep{pat98}. The light curve had a highly significant modulation in the 1966 eruption with a period of 0.099096 days found by \citet{sch90}. Other transient periodicities at 0.1098 days and 1.24 days were found by \citet{pat98} in quiescence. \citet{pat98} suggested that the 1.24-day period might arise from precession in the accretion disc as it is roughly in the range of that in CVs while there is no explanation for the 0.1098-day period.

The Orbital inclination, $i$, is thought to be low from the spectral profiles and low radial velocity amplitude \citep{uth10}. It has been estimated to be $\sim$6$^{\circ}$ \citep{shah97}, 10$^{\circ}$$-$20$^{\circ}$ \citep{pat98}, 20$^{\circ}$$-$30$^{\circ}$ \citep{sel10}, $i$=10$^{\circ}$$\pm$2$^{\circ}$ \citep{uth10} and recently $i$=15$^{\circ}$ \citep{che11}. Such a fairly low inclination is suggested to be the cause of the long dip, which lasts for half of the orbit in the flat top light curve, due to the heating effect on the companion star \citep{pat98}.


T Pyx is the only RN that presents a persistent nova shell. This has a radius of $\sim$5", discovered by \citet{due79}. A fainter outer halo around the shell discovered by \citet{sha97} expands to a radius of $\sim$10". The shell has been found to be slowly expanding with thousands of discrete knots \citep{due79,wil82,sha97} and is suggested to be a result of the shocks from the new eruption ejecta that interact with the ejecta from the previous eruption \citep{con97}.

\begin{figure}[th!]
\begin{center}
\leavevmode
\includegraphics[width=30pc]{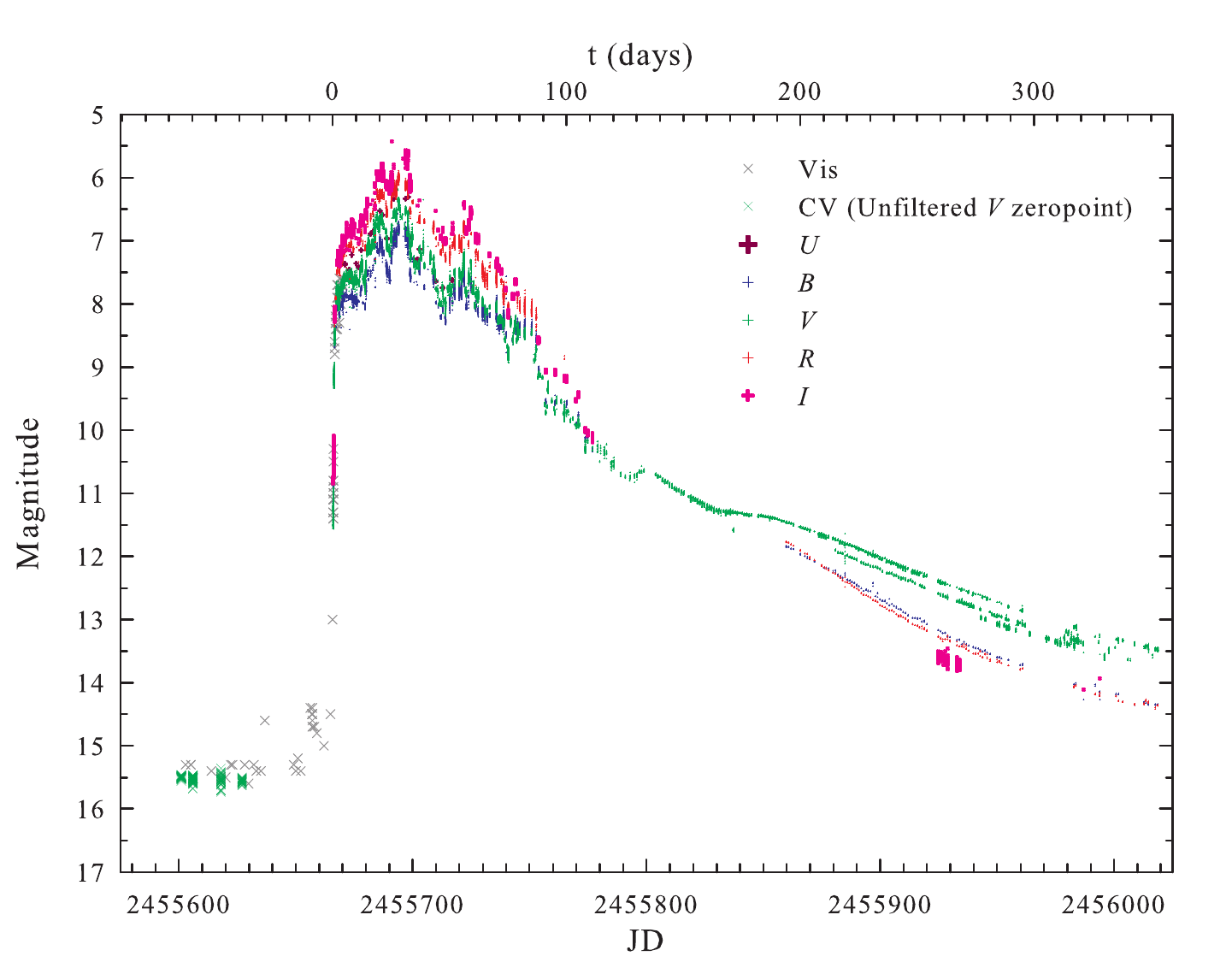} 
\end{center}
\caption[The AAVSO optical light curves of T Pyx in its 2011 outburst.]
{The AAVSO optical light curves of T Pyx in its 2011 outburst.}
\label{tpyx-lc-aavso}
\end{figure}

The outburst of T Pyx in 2011 was discovered by AAVSO observer M. Linnolt at a visual magnitude of 13.0 on 2011 Apr 14.29 UT (JD 2455665.7931, hereafter $t$=0 day) and published in \citet{sch13}. Figure \ref{tpyx-lc-aavso} presents the AAVSO optical light curves of T Pyx in its 2011 outburst from before the eruption to 2012 Mar 31 (JD 2456018, $t$=353 days).

\citet{sch13} reported that from 2008 to 2011 Mar 31 (JD 2455652, $t$=$-$13.8 days) T Pyx's brightness was steady at $V$=15.5 with the usual periodic variation of $<$0.2 mag and it  was never brighter than $V$=15.0. On 2011 Apr 5.51 UT, it was observed with variations from $V$=14.4-14.7 and continued to fade slowly until 2011 Apr 10.54 UT (JD 2455662.0417, $t$=$-$3.7 days).
There was a short brightness increase, a so called ``pre-eruption rise'' (peak around $V$=14.4 mag), occurring about 13 days prior to the initial rise \citep{sch13}. After the sharp initial rise on day $t$=0, a small fading occurred roughly 2 days after that \citep{sch13}.

\citet{pat13} studied the post-outburst light curve in 2011 and found that an orbital period signal appeared by $t$=170 days ($V$=11.2) with a period increase of 0.0054(7)\% implied to be the result of mass ejection from the WD. They derived an ejected mass of at least 3$\times$10$^{-5}$M$_{\odot}$, similar to that in CNe.

Spectroscopic observations of 2011 outburst are discussed in \citet{sho11,sho13} and in \citet{ima12}. \citet{sho11} report high resolution observations at 7 epochs from $t$=1.6-46.6days and \citet{sho13} describe observations at a further 4 epochs much later in the ourburst ($t$=179, 180, 349 and 360 days). \citet{ima12} meanwhile report 11 epochs of low resolution spectroscopy at early times ($t$=2.7-30.8 days). Comparison with results of more extensive and intensive spectroscopic coverage is given below, together with high cadence photometry.

In this paper, in Section 2 we first describe our extensive photometric and spectroscopic datasets. In Section 3 we then use the results of these to explore the various phases of development of the nova both in comparison to those found in other novae and also with models of the outburst. Periodicities in the SMEI light curve are also investigated. Comparison with extensive observations at X-ray and radio wavelengths is detailed in Section 4 and in Section 5 we present our conclusions.


\section{Observations}

\subsection{Photometric observations from SMEI}
Photometric observations of T Pyx were obtained with the Solar Mass Ejection Imager (SMEI). SMEI is an all sky white-light CCD-based camera system for observing the inner heliosphere from Earth orbit, on board the U.S. Air Force $Coriolis$ Mission spacecraft which was launched into an 840 km Sun-synchronous terminator orbit on 2003 Jan 6 \citep{jac04}.
SMEI consists of three baffled CCD cameras each with a roughly 60$^{\circ}$$\times$3$^{\circ}$ field of view.
It continuously takes 4-second exposures, sweeps nearly over all the sky every 102 minutes, and therefore provides about 1500 frames from each camera per Earth orbit \citep{hic05}.

SMEI is operated as a high precision differential photometer \citep{buf06} and can reliably detect brightness changes in point sources down to $m_{{\it SMEI}}$$\sim$8 with the instrument's peak quantum efficiency at $\sim$700 nm and FWHM $\sim$300 nm \citep{hou10}. The SMEI real-time data pipeline produces calibrated all sky maps in which brightness contributions from zodiacal dust and unresolved sidereal background have been subtracted. Approximately 5600 point sources brighter than $m_{{\it SMEI}}$=6 are removed individually from the composite skymaps in order not to compromise the photometric specification for SMEI \citep{hic05}. Therefore with its high-cadence-all-sky observations, SMEI can investigate bright nova explosions ($m_{{\it SMEI}}$$<$8) whose outbursts occurred within the time period of operation (i.e. during 2003-2011) and produce extremely detailed light curves. These caught, for example, the pre-maximum halts during the outbursts of several novae, as presented in \citet{hou10}.

In this work, T Pyx was included in a supplementary star catalog which was added to the UCSD SMEI database of \citet{hou10}. The photometry was produced via the fitting of a modelled PSF \citep{hic07} to the composite skymaps produced by the SMEI data pipeline. The white light observations of T Pyx by SMEI were made from 2011 Apr 15.84 UT (JD 2455667.3422, $t$=1.5 days) to 2011 Jun 2.32 UT (JD 2455714.8165, $t$=49.0 days).

\subsection{Spectroscopic observations}
Spectroscopic observations were obtained with the 1.5m telescope of the Small and Moderate Aperture Research Telescope System (SMARTS) and the 2m Liverpool Telescope (LT). A log of the observations is given in Table \ref{tpyx-spectral-obs-log}.

\begin{table} [tp!]
\begin{center}
\caption[Log of optical spectral observations.]
{Log of optical spectral observations. The time in the third column is counted from the discovery on 2011 Apr 14.29 UT (JD 2455665.7931, $t$=0 days).}
\vspace{0.3cm}
\tiny
\begin{tabular}{ccccccccccccccccccccccccccccccccccccccccccccccccccccccccccccccccccccccccccccccccccccccccccccccccccccccccccccccccccccccccccccccccccccccccc}
\hline \hline
Observation dates & JD	          & $t$ (d) &	 $\lambda$ range (\AA$ $)* & Telescope   & $\vert$ &   Observation dates & JD	          & $t$ (d) &	 $\lambda$ range (\AA$ $)* & Telescope  \\
\hline
2011-04-14 	& 2455666.55056	& 0.8   &			 5630-6950	        & SMARTS   & $\vert$ & 2011-06-23 	 & 2455736.45998	& 70.7  &		 3870-4544		   & SMARTS     \\
2011-04-15 	& 2455667.49308	& 1.7   &				 3250-9520       & SMARTS   & $\vert$ & 2011-06-25 	& 2455738.44247	& 72.6  &				 3250-9520  & SMARTS     \\
2011-04-16 	& 2455668.44853	& 2.7   &	3655-5424			        & SMARTS   & $\vert$ & 2011-06-26 	& 2455739.45381	& 73.7  &		 3870-4544		   & SMARTS     \\
2011-04-17 	& 2455669.43890	& 3.6   &			 5630-6950	        & SMARTS   & $\vert$ & 2011-06-28 	 & 2455741.40939	& 75.6  &		 3870-4544		   & SMARTS     \\
2011-04-18 	& 2455670.51676	& 4.7   &		3870-4544		        & SMARTS   & $\vert$ & 2011-06-30 	& 2455743.44596	& 77.7  &	 3655-5424			   & SMARTS     \\
2011-04-19 	& 2455671.47293	& 5.7   &			 5630-6950	        & SMARTS   & $\vert$ & 2011-07-01 	 & 2455744.47839	& 78.7  &				 3250-9520  & SMARTS     \\
2011-04-22 	& 2455674.38198	& 8.6 	& 3900-5700, 5800-9400	& LT       & $\vert$ & 2011-07-02 	& 2455745.46245	& 79.7  &		 3870-4544		   & SMARTS     \\
2011-04-22* 	& 2455674.61188	& 8.8 &			 5630-6950	        & SMARTS   & $\vert$ & 2011-07-03 	 & 2455746.49198	& 80.7  &			 5630-6950	   & SMARTS     \\
2011-04-23 	& 2455675.38192	& 9.6 	& 3900-5700, 5800-9400  & LT       & $\vert$ & 2011-09-15  & 2455820.92030	& 155.1 &	 3655-5424			   & SMARTS     \\
2011-04-23 	& 2455675.46047	& 9.7   &	3655-5424			        & SMARTS   & $\vert$ & 2011-09-19  & 2455824.89457	& 159.1 &	 3655-5424			   & SMARTS     \\
2011-04-24 	& 2455676.38186	& 10.6	& 3900-5700, 5800-9400  & LT       & $\vert$ & 2011-09-21  & 2455826.89632	& 161.1 &			 5630-6950	   & SMARTS     \\
2011-04-24 	& 2455676.50600	& 10.7  &			 5630-6950	        & SMARTS   & $\vert$ & 2011-09-21a & 2455826.92193	& 161.1 &		 4060-4735		   & SMARTS     \\
2011-04-25 	& 2455677.38180	& 11.6	& 3900-5700, 5800-9400  & LT       & $\vert$ & 2011-09-23  & 2455828.91213	& 163.1 &			 3250-9520	   & SMARTS     \\
2011-04-25 	& 2455677.44188	& 11.6  &	3655-5424			        & SMARTS   & $\vert$ & 2011-09-25  & 2455830.87711	& 165.1 &	 3655-5424			   & SMARTS     \\
2011-04-27 	& 2455679.45705	& 13.7  &	3655-5424			        & SMARTS   & $\vert$ & 2011-09-27  & 2455832.89575	& 167.1 &			 5630-6950	   & SMARTS     \\
2011-04-28 	& 2455680.55124	& 14.8  &			 5630-6950	        & SMARTS   & $\vert$ & 2011-09-29  & 2455834.86413	& 169.1 &		 3870-4544		   & SMARTS     \\
2011-04-29 	& 2455681.38713	& 15.6  &			 5630-6950	        & SMARTS   & $\vert$ & 2011-10-03  & 2455838.86920	& 173.1 &		 3870-4544		   & SMARTS     \\
2011-05-01 	& 2455683.51565	& 17.7  &				 3250-9520       & SMARTS   & $\vert$ & 2011-10-03a & 2455838.89211	& 173.1 &			 5630-6950	   & SMARTS     \\
2011-05-02 	& 2455684.46054	& 18.7  &	3655-5424			        & SMARTS   & $\vert$ & 2011-10-05  & 2455840.83755	& 175.0 &		 3870-4544		   & SMARTS     \\
2011-05-03 	& 2455685.29780	& 19.5  &		3870-4544		        & SMARTS   & $\vert$ & 2011-10-05a & 2455840.87123	& 175.1 &			 5630-6950	   & SMARTS     \\
2011-05-04 	& 2455686.44255	& 20.6  &		3870-4544		        & SMARTS   & $\vert$ & 2011-10-07  & 2455842.85832	& 177.1 &				 3250-9520  & SMARTS     \\
2011-05-05 	& 2455687.45694	& 21.7  &	3655-5424			        & SMARTS   & $\vert$ & 2011-10-09  & 2455844.85722	& 179.1 &			 5630-6950	   & SMARTS     \\
2011-05-07 	& 2455689.46274	& 23.7  &	3655-5424			        & SMARTS   & $\vert$ & 2011-10-15  & 2455850.84231	& 185.0 &			 5630-6950	   & SMARTS     \\
2011-05-08 	& 2455690.45846	& 24.7  &			 6230-7550	        & SMARTS   & $\vert$ & 2011-10-16  & 2455851.86406	& 186.1 &	 		  3250-9520  & SMARTS     \\
2011-05-09 	& 2455691.47495	& 25.7  &	3655-5424			        & SMARTS   & $\vert$ & 2011-10-17  & 2455852.86060	& 187.1 &			 5630-6950	   & SMARTS     \\
2011-05-10 	& 2455692.45741	& 26.7  &			 5630-6950	        & SMARTS   & $\vert$ & 2011-10-19  & 2455854.83327	& 189.0 &	 3655-5424			   & SMARTS     \\
2011-05-11 	& 2455693.46645	& 27.7  &	3655-5424			        & SMARTS   & $\vert$ & 2011-10-23  & 2455858.83362	& 193.0 &			 5630-6950	   & SMARTS     \\
2011-05-13 	& 2455695.45393	& 29.7  &	3655-5424			        & SMARTS   & $\vert$ & 2011-10-29  & 2455864.78561	& 199.0 &	 3655-5424			   & SMARTS     \\
2011-05-14 	& 2455696.55773	& 30.8  &		3870-4544		        & SMARTS   & $\vert$ & 2011-11-01  & 2455867.83583	& 202.0 &				 3250-9520  & SMARTS     \\
2011-05-15 	& 2455697.45702	& 31.7  &				 3250-9520       & SMARTS   & $\vert$ & 2011-11-03  & 2455869.82510	& 204.0 &			 5630-6950	   & SMARTS     \\
2011-05-16 	& 2455698.43956	& 32.6  &			 5630-6950	        & SMARTS   & $\vert$ & 2011-11-05  & 2455871.82328	& 206.0 &	 3655-5424			   & SMARTS     \\
2011-05-17 	& 2455699.46366	& 33.7  &			 5985-9480	        & SMARTS   & $\vert$ & 2011-11-15  & 2455881.78178	& 216.0 &		 3870-4544		   & SMARTS     \\
2011-05-18 	& 2455700.40413	& 34.6  &	3655-5424			        & SMARTS   & $\vert$ & 2011-11-15a & 2455881.80037	& 216.0 &			 5630-6950	   & SMARTS     \\
2011-05-20 	& 2455702.45740	& 36.7  &			 5630-6950	        & SMARTS   & $\vert$ & 2011-11-16  & 2455882.77670	& 217.0 &				 3250-9520  & SMARTS     \\
2011-05-24 	& 2455706.43767	& 40.6  &	3655-5424			        & SMARTS   & $\vert$ & 2011-11-19  & 2455885.78113	& 220.0 &	 3655-5424			   & SMARTS     \\
2011-05-25 	& 2455707.45922	& 41.7  &	3655-5424			        & SMARTS   & $\vert$ & 2011-11-21  & 2455887.74116	& 221.9 &		 3870-4544		   & SMARTS     \\
2011-05-26 	& 2455708.54164	& 42.7  &		3870-4544		        & SMARTS   & $\vert$ & 2011-11-21a & 2455887.84304	& 222.0 &			 5630-6950	   & SMARTS     \\
2011-05-29 	& 2455711.37204	& 45.6  &	3655-5424			        & SMARTS   & $\vert$ & 2011-11-23  & 2455889.74104	& 223.9 &				 3250-9520  & SMARTS     \\
2011-05-30 	& 2455712.45091	& 46.7  &			 5630-6950	        & SMARTS   & $\vert$ & 2011-11-25  & 2455891.80641	& 226.0 &			 5630-6950	   & SMARTS     \\
2011-05-31 	& 2455713.45613	& 47.7  &			 5630-6950	        & SMARTS   & $\vert$ & 2011-11-29  & 2455895.69458	& 229.9 &	 3655-5424			   & SMARTS     \\
2011-06-01 	& 2455714.43748	& 48.6  &				 3250-9520       & SMARTS   & $\vert$ & 2011-12-08  & 2455904.67796	& 238.9 &	 3655-5424			   & SMARTS     \\
2011-06-02 	& 2455715.50471	& 49.7  &			 5630-6950	        & SMARTS   & $\vert$ & 2011-12-09  & 2455905.69137	& 239.9 &		 3870-4544		   & SMARTS     \\
2011-06-03 	& 2455715.51736 & 49.7  &	3655-5424			        & SMARTS   & $\vert$ & 2011-12-12  & 2455908.67536	& 242.9 &	 3655-5424			   & SMARTS     \\
2011-06-08 	& 2455721.51827	& 55.7  &	3655-5424			        & SMARTS   & $\vert$ & 2011-12-13  & 2455909.81406	& 244.0 &			 5630-6950	   & SMARTS     \\
2011-06-09 	& 2455722.45866	& 56.7  &		3870-4544		        & SMARTS   & $\vert$ & 2011-12-14  & 2455910.66410	& 244.9 &		 3870-4544		   & SMARTS     \\
2011-06-10 	& 2455723.47018	& 57.7  &				 3250-9520       & SMARTS   & $\vert$ & 2011-12-14a & 2455910.77023	& 245.0 &			 5630-6950	   & SMARTS     \\
2011-06-11 	& 2455724.45218	& 58.7  &	3655-5424			        & SMARTS   & $\vert$ & 2011-12-15  & 2455911.76083	& 246.0 &	 3655-5424			   & SMARTS     \\
2011-06-12 	& 2455725.44831	& 59.7  &			 5630-8860	        & SMARTS   & $\vert$ & 2011-12-16  & 2455912.74787	& 247.0 &				 3250-9520  & SMARTS     \\
2011-06-14 	& 2455727.47669	& 61.7  &			 5630-6950	        & SMARTS   & $\vert$ & 2011-12-17  & 2455913.74309	& 247.9 &			 5630-6950	   & SMARTS     \\
2011-06-16 	& 2455729.46994	& 63.7  &		5630-6950           & SMARTS   & $\vert$ & 2011-12-18  & 2455914.73296	& 248.9 &		 3870-4544		   & SMARTS     \\
2011-06-22 	& 2455735.45566	& 69.7  &	3655-5424			        & SMARTS   & $\vert$ & 2011-12-19  & 2455915.72555	& 249.9 &			 5630-6950	   & SMARTS     \\
\hline \hline
\label{tpyx-spectral-obs-log}
\end{tabular}
\end{center}
\vspace{-0.7cm}
\small
\begin{center}
* Spectra with $\lambda$ range 3250-9518\AA$ $ are taken with low resolution.
\end{center}
\end{table}

\subsubsection{Small and Moderate Aperture Research Telescope System (SMARTS)}
The 1.5m telescope of the SMARTS II Consortium located at CTIO Chile is equipped with a long-slit R-C spectrograph. An ultraviolet transmitting grating is used at the f/7.5 focus with a plate scale 18.1 arcseconds/mm and a 1200$\times$800 CCD. We obtained 99 low-to-moderate-resolution (300 $<$ R $<$ 3400) optical spectra of T Pyx. These spectra include 49 moderate-resolution spectra in the blue region (3655-5424\AA$ $ and 3870-4544\AA$ $), 37 moderate-resolution spectra in the red region (5630-6950\AA$ $), and 13 low-resolution spectra in a wide wavelength region (3250-9520\AA$ $). 56 spectra of particular interest here were obtained from 2011 Apr 14 (JD 2455666.55056, $t$=0.757 days, given hereafter 0.8 days), which was the earliest epoch T Pyx was observed spectroscopically, to 2011 Jul 3 (JD 2455746.49198, $t$=80 days). Another 43 spectra were obtained from 2011 Sep 15 (JD 2455820.92030, $t$=155.127) to 2011 Dec 19 (JD 2455915.72555, $t$=249.9 days). All data are available through the SMARTS atlas - see \citet{wal12}\footnote{Available online at http://www.astro.sunysb.edu/fwalter/SMARTS/NovaAtlas/tpyx/tpyx.html}.

\subsubsection{Liverpool Telescope (LT)}
The 2m robotic Liverpool Telescope (LT; \citealp[see][]{ste04}) is sited at the Observatorio del Roque de Los Muchachos on the Canary Island of La Palma, Spain. T Pyx's spectra on the nights of 2011 Apr 22-25 were secured using the Fibre-fed RObotic Dual-beam Optical Spectrograph (FRODOSpec; \citealp[see][]{bar12}) on LT over 3900-5700\AA$ $ in the blue (R$\sim$2200) and 5800-9400\AA$ $ in the red arm (R$\sim$2600), with exposures of 60 s. Data reduction was performed through a pipeline that initially performs bias, dark frame, and flat field subtraction. A spectroscopic standard star HD289002 ($\alpha$=06$^{h}$45$^{m}$13$^{s}\!.$371, $\delta$=+02$^{\circ}$08$'$14$''\!.$70) was observed at a similar airmass and used to remove instrumental and atmospheric response. We used the $onedspec$ package in IRAF\footnote{IRAF is distributed by the National Optical Astronomy Observatories, which are operated by the Association of Universities for Research in Astronomy, Inc., under cooperative agreement with the National
Science Foundation.} to analyse all LT spectra.

\section{Results and Discussion}

\subsection{SMEI light curves}
\begin{figure}[bph!]
\begin{center}
\leavevmode
\epsfxsize = 14.0cm
\epsfysize = 14.0cm
\includegraphics[width=30pc]{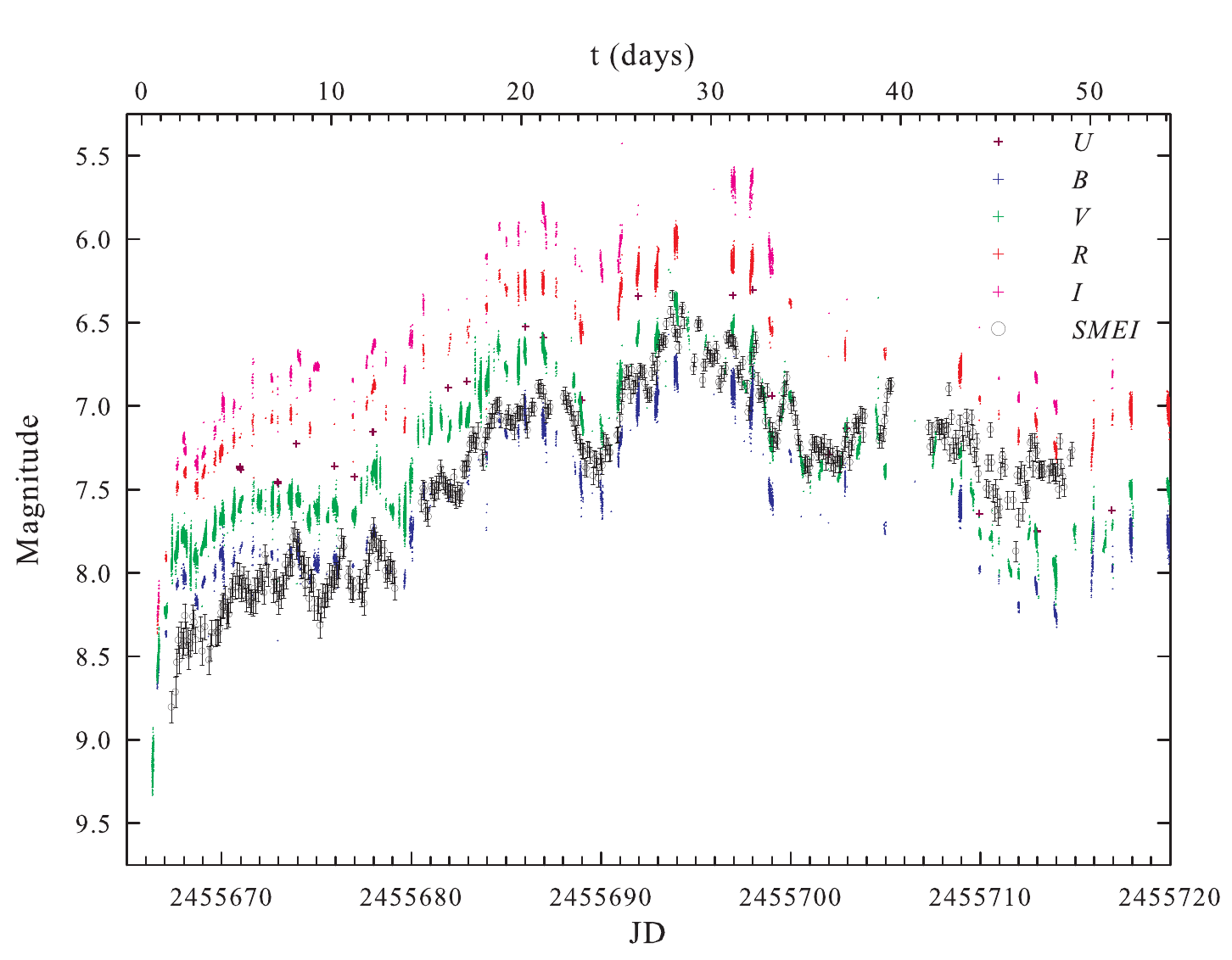} 
\end{center}
\caption[The SMEI light curve of T Pyx during its 2011 outburst.]
{The SMEI light curve of T Pyx during its 2011 outburst (open black circles with error bars) compared to $UBVRI$ light curves observed by AAVSO (plus signs).}
\label{tpyx-smei-lc}
\end{figure}

\begin{figure}[tph!]
\begin{center}
\leavevmode
\epsfxsize = 14.0cm
\epsfysize = 14.0cm
\includegraphics[width=25pc]{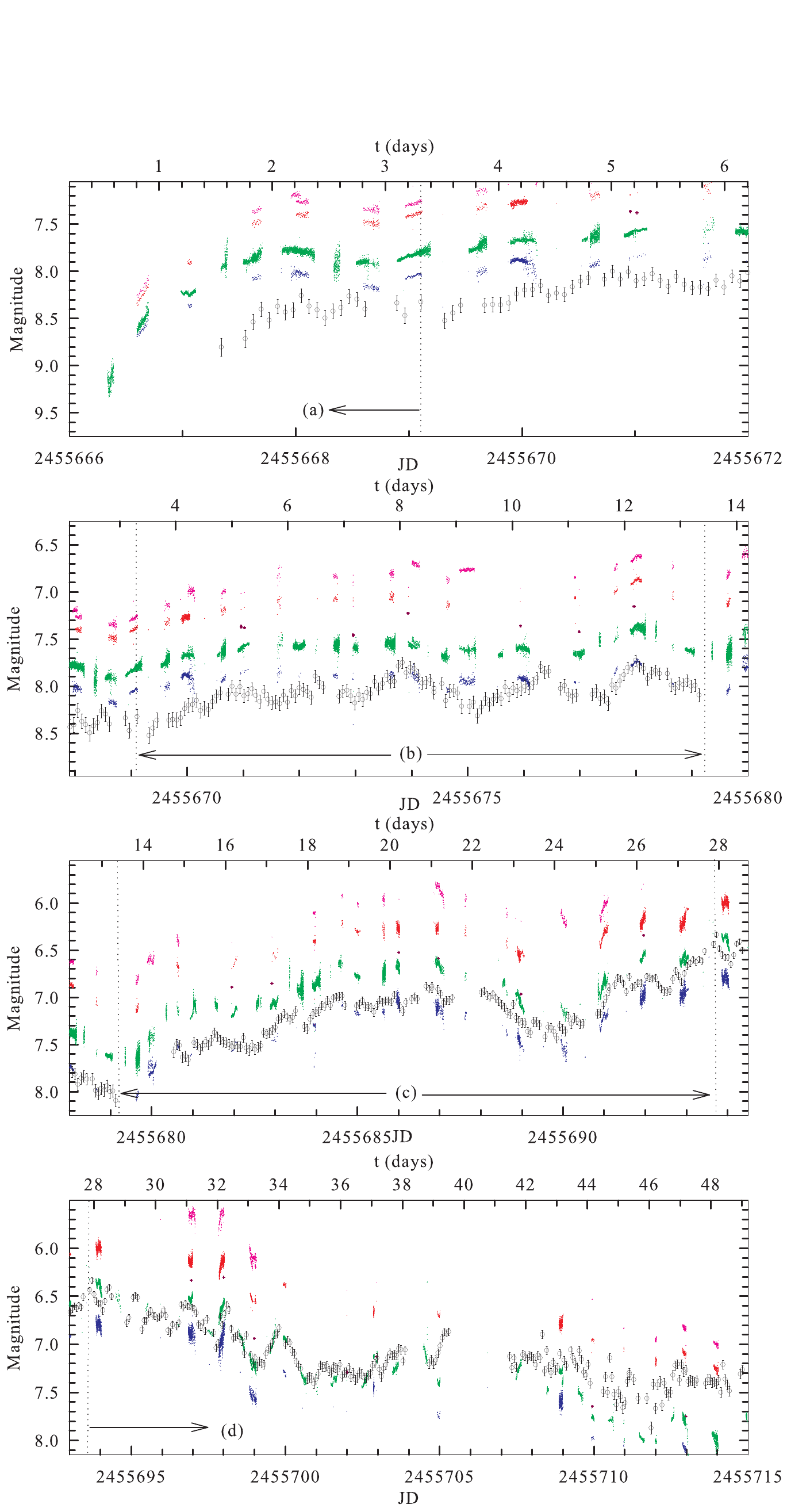} 
\end{center}
\caption[Four idealised phases of the SMEI light curve.]
{Four idealised phases of the SMEI light curves including the initial rise (a), the pre-maximum halt (b), the final rise (c), and the early decline (d). Vertical dotted lines separate each phase of the light curve.}
\label{tpyx-smei-lc-4panels}
\end{figure}

The SMEI light curve is compiled from 533 observations\footnote{Available online data of SMEI sky map at http://smei.ucsd.edu/sky/index.html} and provides unprecedented detail with high cadence data that are compared to AAVSO light curves in Figures \ref{tpyx-smei-lc} and \ref{tpyx-smei-lc-4panels}.

\subsubsection{Phases of the Outburst from the SMEI light curve}

We divide the SMEI light curve into 4 parts based on the idealised nova optical light curve given in \citet{war08} - see Figure \ref{tpyx-smei-lc-4panels}:
\begin{enumerate}
\item {\itshape The initial rise (2011 Apr 15-16, $t$=1.5-3.3 days):\/} The first reliable detection of the nova outburst by SMEI occurred at the end of the rapid rise seen in AAVSO data, at $m_{{\it SMEI}}$=8.80$\pm$0.09 on 2011 Apr 15.84 UT (JD 2455667.3422, $t$=1.5 days). After that time SMEI observed T Pyx approximately every 102 minutes. The light curve rose to $m_{{\it SMEI}}$=8.26 on JD 2455668.0476 ($t$=2.2 days) and began quasi-periodic variations \citep{hou11b} as shown in the top left panel of Figure \ref{tpyx-smei-lc-4panels}. This phase ended at $t$$\sim$3.3 days when the light curve began to flatten.

\item {\itshape Pre-maximum halt (2011 Apr 16-27, $t$=3.3-13.3 days):\/} Following the initial rise, the SMEI light curve rose very slowly (noted as ``almost plateau'', by \citet{hou11b}) with clear quasi-periodic variations as shown in Figure \ref{tpyx-smei-lc-4panels} (b). The peak-to-peak times of the variations vary between 1.1-2.8 days. The magnitudes ranged between 7.7-8.5 mags. The first peak and dip of this variation was also noticed by \citet{sch13}. However the AAVSO light curve did not clearly show the quasi-periodic variations detectable in the SMEI light curve. This phase may coincide with the pre-maximum halt defined as a pause at about 2 magnitudes below maximum for a few days for slow novae \citep{war08,hou10}. There was then a gap in the SMEI data (due to instrumental problems) lasting 1.4 days ($t$=13.3-14.7 days).

\item {\itshape Final rise (2011 Apr 28 - May 11, $t$=14.7-27.9 days):\/} After the pre-maximum halt phase, the light curve was seen to rise more steeply to $m_{{\it SMEI}}$=6.88$\pm$0.04 at $t$=20.9 days. Meanwhile the apparent quasi-periodic variations still persisted with approximately half the amplitude of those seen in the previous phase.

    The light curve then suddenly dropped at $t$=22.1 days to $m_{{\it SMEI}}$$\sim$7.4 for about two days. This event appeared as a major dip in the light curve before it reached maximum light. After this dip, the light curve rose again to $m_{{\it SMEI}}$=6.79$\pm$0.04 at $t$=25.5 days and stayed there for roughly two days before it reached visual maximum at $m_{{\it SMEI}}$=6.33$\pm$0.03 on 2011 May 12.22 (JD 2455693.7251, $t$=27.9 days) as seen in the bottom left panel of Figure \ref{tpyx-smei-lc-4panels}.

\item {\itshape Early decline (2011 May 11 - - [$\sim$Oct 3] , $t$=27.9 days - - [$\sim$90 days]):\/} After optical maximum, the nova declined with variations (amplitude ranges roughly between 0.1-0.3 mags) to $m_{{\it SMEI}}$$\sim$7.2-7.4 from 2011 May 19.42-20.76 UT (JD 2455700.9203-2455702.2605, $t$=35.1-36.5 days). The light curve experienced a dip again at around $t$$\sim$ 36 days and followed this with a broad hump lasting from $t$$\sim$ 44-47 days as seen in Figure \ref{tpyx-smei-lc-4panels} (d). The last reliable detection of T Pyx by SMEI was at $m_{{\it SMEI}}$=7.26$\pm$0.05 on 2011 Jun 2.37 UT (JD 2455714.8165, $t$=49.02 days).

    Further light curve observations provided by the AAVSO in Figure \ref{tpyx-lc-aavso} show that the early decline phase ends at around $t$$\sim$90 days with the brightness having declined by approximately 2.5-3.5 magnitudes from maximum. The final decline ($\sim$6 magnitudes from maximum) began around $t$$\sim$260 days.
\end{enumerate}

We may compare the early stages of the outburst captured by SMEI to the light curves presented by \citet{hil13} from TNR modelling with a range of parameters. There is certainly some resemblance to some of these model light curves in terms of the pre-maximum halt and subsequent smaller reversal just before optical peak. The former is attributed to a temporary drop in energy flux as convection in the expanding, thinning envelope ceases to be efficient near the surface of the envelope. The later dip resembles the shorter time scale feature noticed in e.g. RS Oph and KT Eri by \citet{hou10} just before maximum light.

When directly comparing the SMEI to the AAVSO light curves, we see that the nova seems to be bluer and similar to the magnitude in the $B$ filter at the pre-maximum halt phase, then exhibits the same brightness in the $B$ and $V$ filters at visual maximum where it is expected to behave like an early type star, usually in the range from B5-F5 \citep{war08}. After visual maximum, it tends to be redder, approaching the magnitude in the $R$ filter as shown in Figure \ref{tpyx-smei-lc}.

\begin{figure}[th!]
\begin{center}
\leavevmode
\epsfxsize = 14.0cm
\epsfysize = 14.0cm
\includegraphics[width=30pc]{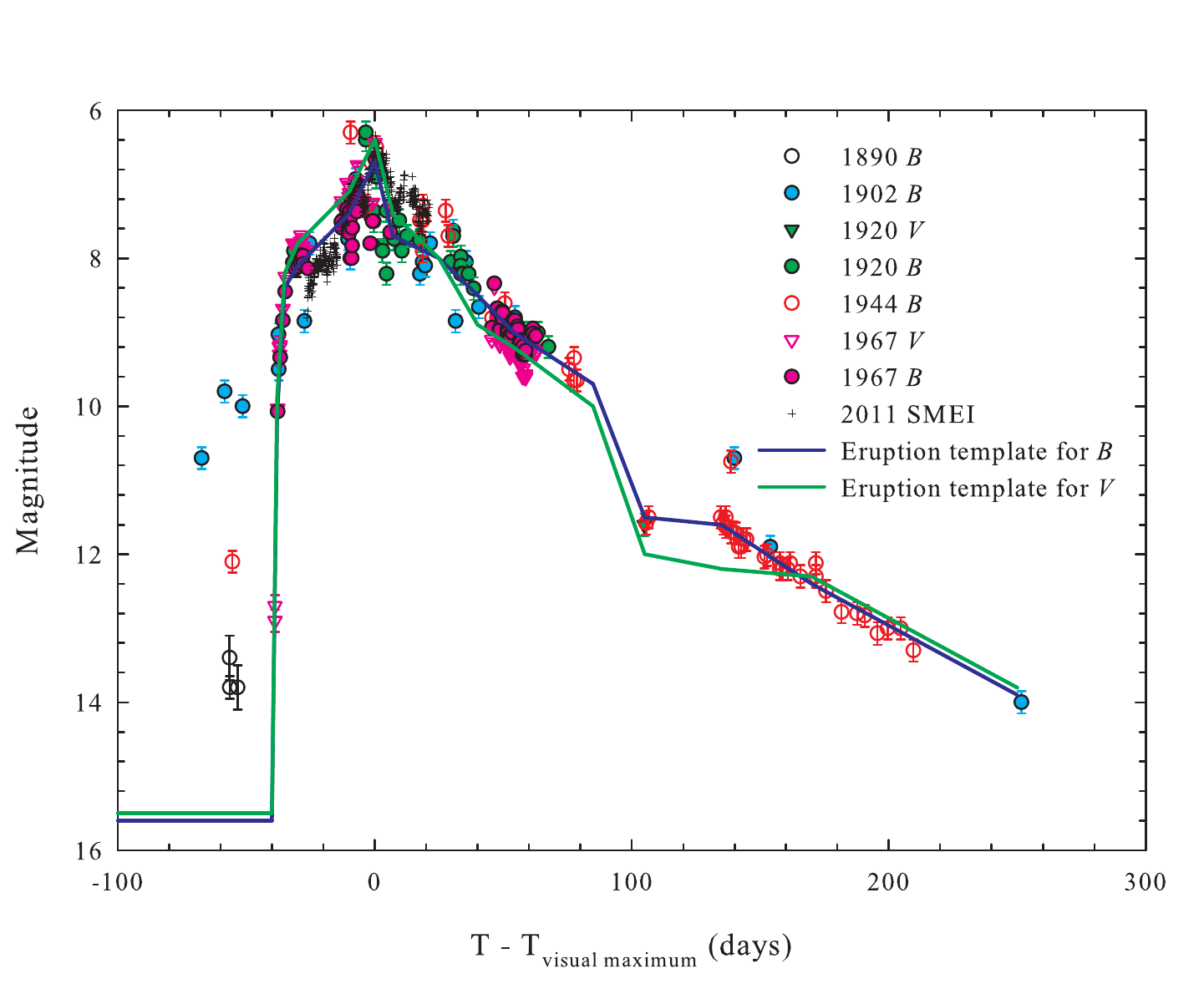} 
\end{center}
\caption[The outburst in 2011 from SMEI compared to previous outbursts.]
{The outburst in 2011 from SMEI (plus signs) compared to previous outbursts. The $B$ (circles) and $V$ (triangles) observations are highlighted in different colours to represent each year of outburst (1890:open black, 1902:blue, 1920:green, 1944:open red, 1967:pink). Data for the previous outbursts are taken from \citet{sch10b}.}
\label{tpyx-compare-previous-outbursts}
\end{figure}

When comparing the SMEI data to the previous five outbursts in $B$ and $V$ and their resulting eruption templates provided in \citet{sch10b}, we find that the shape and the magnitude of the visual peak in 2011 are compatible with previous outbursts. However the pre-maximum halt phase in 2011 seems to be fainter while during the decline after the visual maximum the nova is brighter in 2011 than shown in the templates (Figure \ref{tpyx-compare-previous-outbursts}). All the maxima, including the 2011 maximum, have exhibited brightness fluctuations near the optical peak and the declines have been slow. This is consistent with observations noted in \citet{cat69} and \citet{wil82} for the 1966 outburst.

\subsubsection{Investigation of Periodicities}
The SMEI light curve was searched for any periodic modulations. The analysis was undertaken using all the available SMEI data but we separated it into 4 cases which are (a) from the first observation to the last observation, (b) from the first observation to visual maximum, (c) from visual maximum to the last observation, and (d) from the first observation to the end of the pre-maximum halt phase. The analysis used the $PERIOD04$\footnote{Available online at http://www.univie.ac.at/tops/Period04/} PC code from \citet{len04,len05}. $PERIOD04$ is especially useful for analysis of astronomical time series containing gaps. The program extracts the individual frequencies from the multi-periodic content and provides the frequencies, semi-amplitude, and phase of the harmonic signals of the light curve by using a combination of least-squares fitting and the discrete Fourier transform algorithm. The uncertainty of the estimated periods was derived from Monte Carlo simulations.

\begin{figure}[tph!]
\begin{center}
\leavevmode
\epsfxsize = 14.0cm
\epsfysize = 14.0cm
\includegraphics[width=27pc]{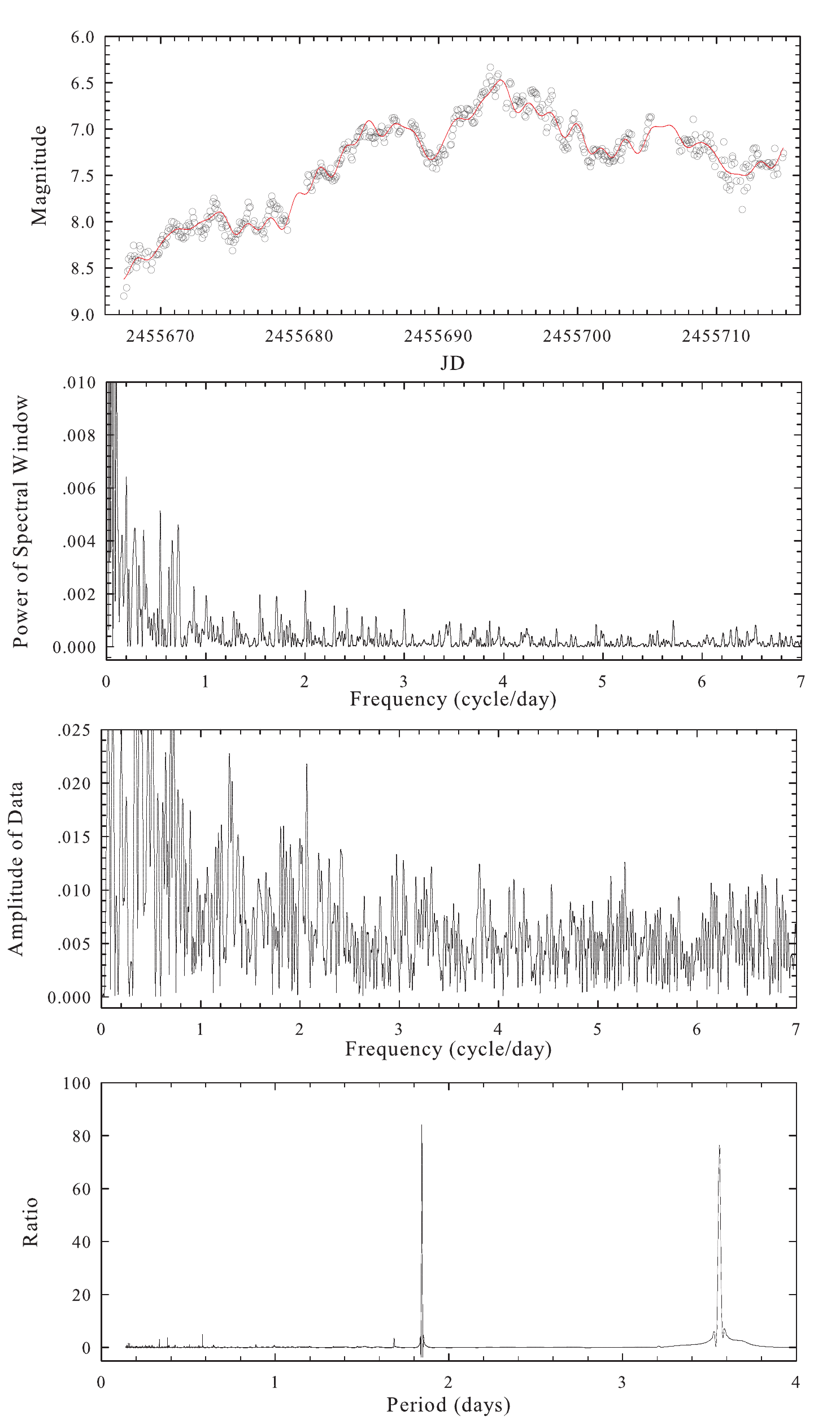} 
\end{center}
\caption[Period found by analysing the SMEI light curve from the beginning of SMEI observations to the last observation.]
{Period found by analysing the SMEI light curve from the beginning of SMEI observations to the last observation. The observation points were fitted (top) and yielded the spectral window (below), the observational data (middle), and the ratio of the spectral window to the amplitude which gives us the possible periods (bottom).}
\label{tpyx-period-begin-end}
\end{figure}

\begin{figure}[tph!]
\begin{center}
\leavevmode
\epsfxsize = 14.0cm
\epsfysize = 14.0cm
\includegraphics[width=27pc]{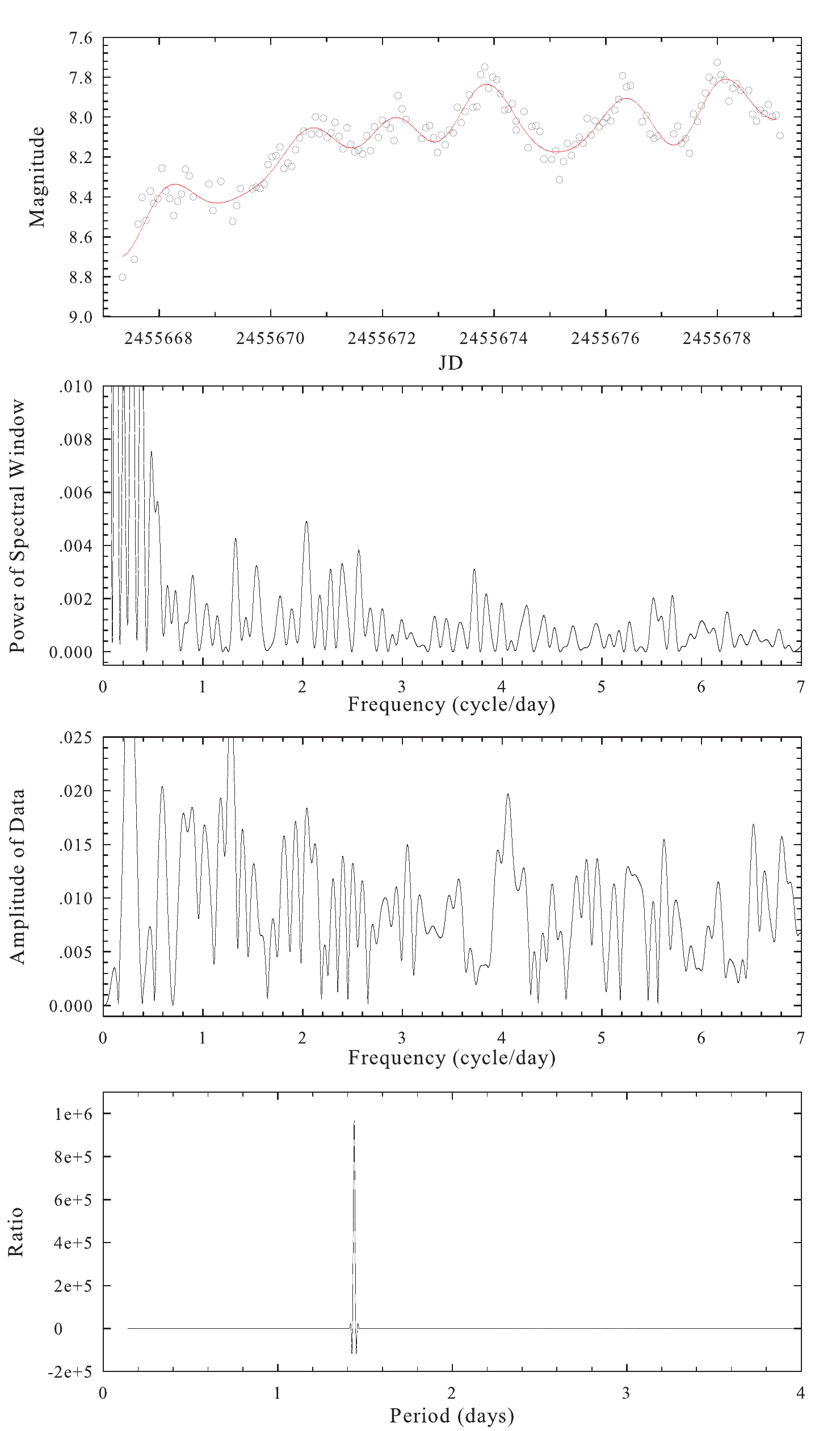} 
\end{center}
\caption[Period found by analysing the SMEI light curve from the beginning of SMEI observations to the end of the pre-maximum halt phase.]
{As Figure \ref{tpyx-period-begin-end} but from the beginning of SMEI observations to the end of the pre-maximum halt phase.}
\label{tpyx-period-begin-gap6}
\end{figure}

\begin{figure}[tph!]
\begin{center}
\leavevmode
\epsfxsize = 14.0cm
\epsfysize = 14.0cm
\includegraphics[width=27pc]{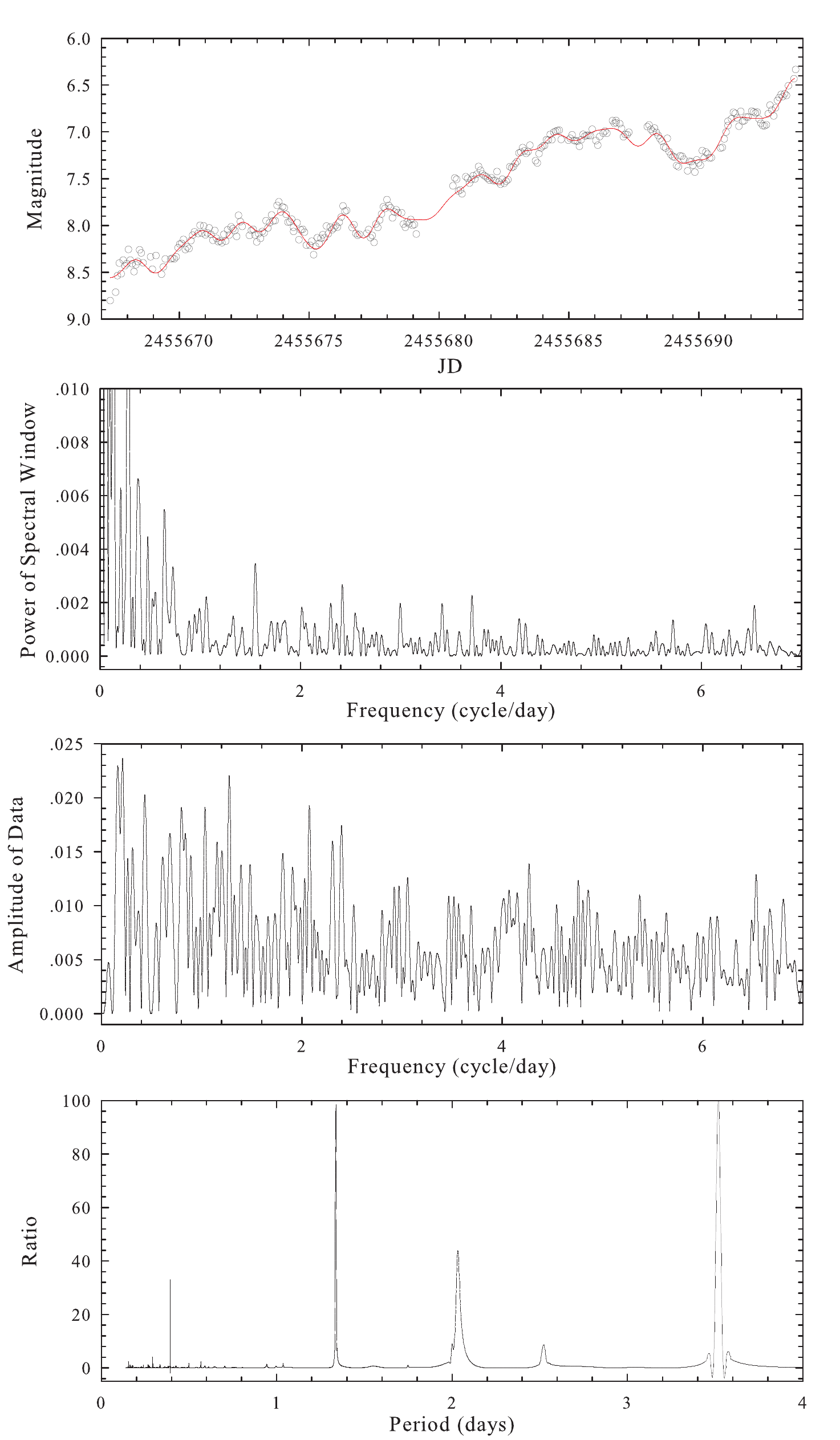} 
\end{center}
\caption[Period found by analysing the SMEI light curve from beginning of SMEI observations to visual maximum.]
{As Figure \ref{tpyx-period-begin-end} but from the beginning of SMEI observations to visual maximum.}
\label{tpyx-period-begin-max}
\end{figure}

\begin{figure}[tph!]
\begin{center}
\leavevmode
\epsfxsize = 14.0cm
\epsfysize = 14.0cm
\includegraphics[width=27pc]{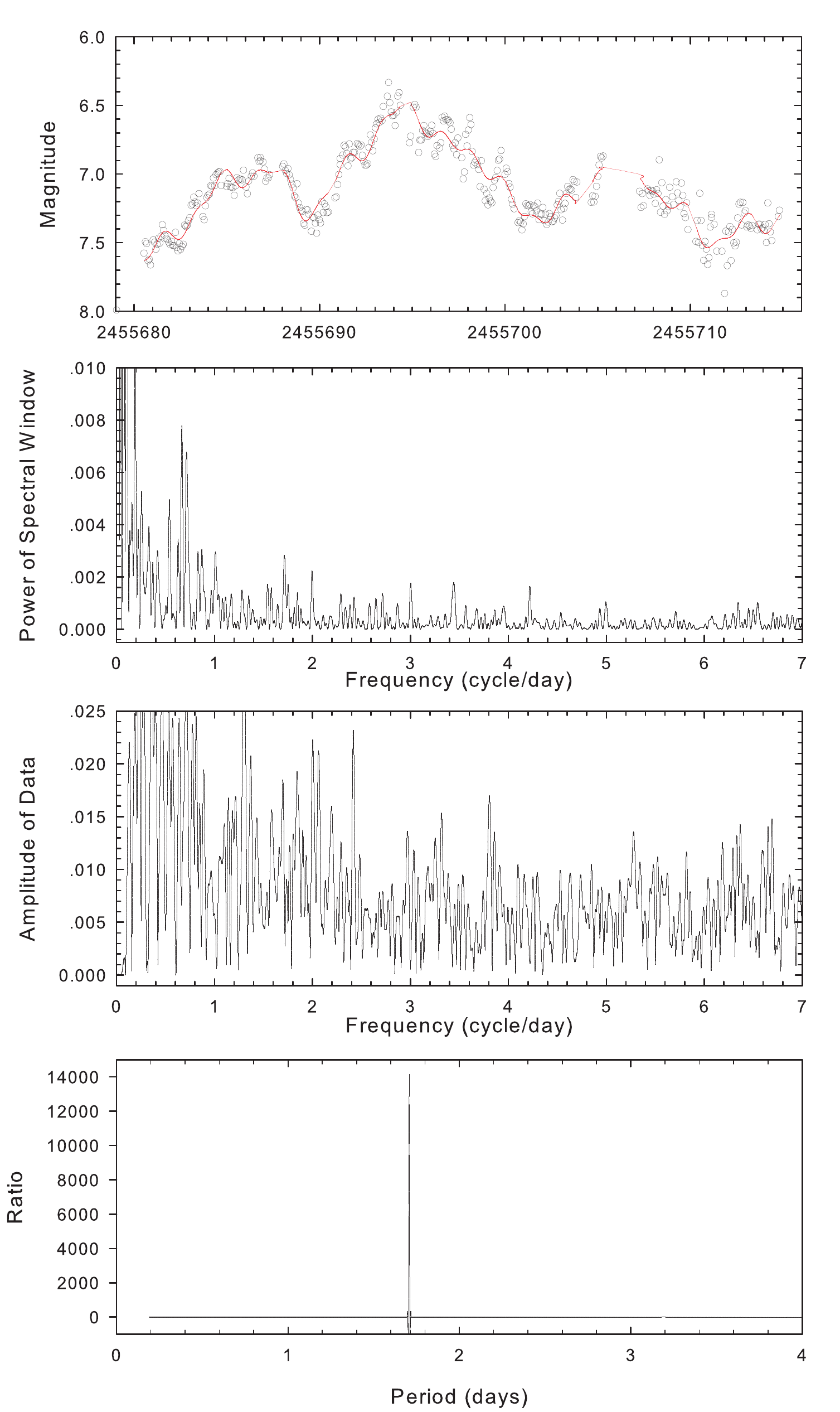} 
\end{center}
\caption[Period found by analysing the SMEI light curve from the end of the pre-maximum halt phase to the last SMEI observation.]
{As Figure \ref{tpyx-period-begin-end} but from the end of the pre-maximum halt phase to the last SMEI observation.}
\label{tpyx-period-gap-end}
\end{figure}

\begin{figure}[tph!]
\begin{center}
\leavevmode
\epsfxsize = 14.0cm
\epsfysize = 14.0cm
\includegraphics[width=27pc]{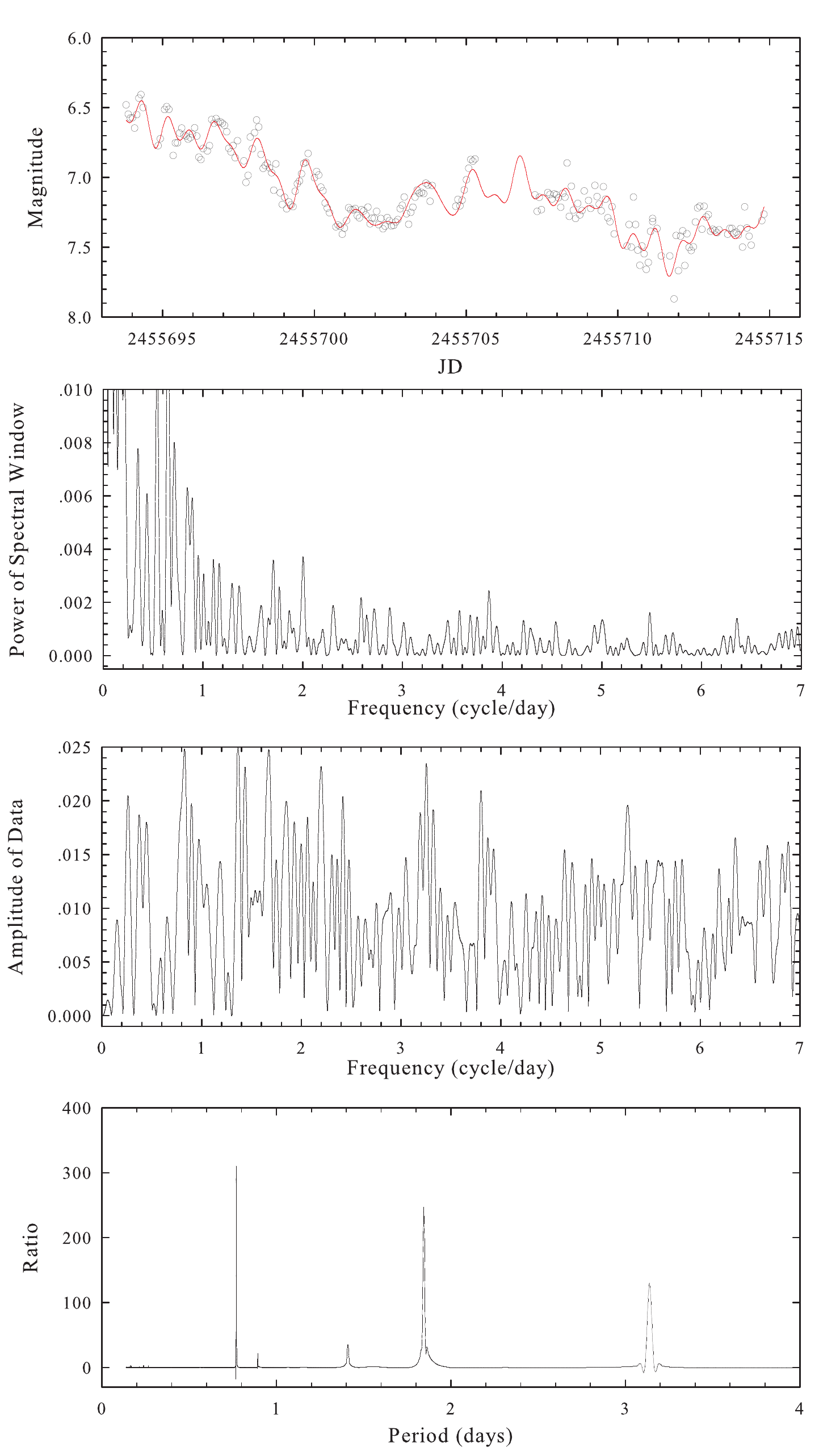} 
\end{center}
\caption[Period found by analysing the SMEI light curve from the visual maximum to the last SMEI observation.]
{As Figure \ref{tpyx-period-begin-end} but from the visual maximum to the last SMEI observation.}
\label{tpyx-period-max-end}
\end{figure}

The resulting periodograms for all 5 cases are displayed in Figures \ref{tpyx-period-begin-end}, \ref{tpyx-period-begin-gap6}, \ref{tpyx-period-begin-max}, \ref{tpyx-period-gap-end}, and \ref{tpyx-period-max-end}. The top panels show how closely the calculations agree with the observations, and the most prominent peaks are presented in Table \ref{tpyx-periods}. A prominent period of 1.8 days is apparent after the pre-maximum halt. However, by far the most strongly detected period, $P$=1.44$\pm$0.05 days, with the highest signal ratio $\sim$10$^{6}$ as shown in Figure \ref{tpyx-period-begin-gap6}, is found up to this time. This period is close to the weak signal of 1.24 days found by \citet{pat98} who suggested it might originate from precession in an accretion disc. However, it is doubtful whether the disc would re-form so soon after the outburst, and at this early time, the central system is expected to lie well within the pseudo-photosphere (see below) and therefore not be directly observed.

\begin{table}[pbh!]
\begin{center}
\caption[Results from $PERIOD04$]
{Results from $PERIOD04$}
\vspace{0.3cm}
\small
\begin{tabular}{c l c c}
\hline \hline
Cases & Parts of light curve                & $t$ (days) & Most prominent period (days)        \\
\hline
(a) &  first observation - last observation &  1.5-49.0  & 1.8$\pm$0.1, 3.6$\pm$0.5    \\
(b) &  first observation - end of pre-maximum halt   &  1.5-13.4  & 1.44$\pm$0.05 \\
(c) &  first observation - visual maximum   &  1.5-27.9  & 3.5$\pm$0.08, 1.34$\pm$0.03   \\
(d) &  end of pre-maximum halt - last observation    &  13.4-49.0 & 1.7$\pm$0.1 \\
(e) &  visual maximum - last observation    &  27.9-49.0 & 0.77$\pm$0.02, 1.84$\pm$0.05 \\
\hline \hline
\label{tpyx-periods}
\end{tabular}
\end{center}
\vspace{-0.7cm}
\small
\begin{center}
\end{center}
\end{table}

Some of the TNR models by \citet{hil13} produced marked oscillations prior to or during extensive mass loss, caused by the restructuring and rebalancing of the envelope as it expands. However, we found no significant changes in H$\alpha$ line shape with the 1.44 day period when we compared the four red spectra ($t$=5.7, 8.6, 9.6, 11.6 days) taken during the optical pre-maximum halt. Similarly, we found no significant differences in the overall ionization during a cycle from consideration of the appearance of lines across the full spectral range. These two aspects are also found to be the same in the blue spectra, where the H$\gamma$ line was of course investigated rather than H$\alpha$.

\subsection{Spectra from SMARTS and LT}
\begin{figure}[phb!]
\begin{center}
\leavevmode
\epsfxsize = 14.0cm
\epsfysize = 14.0cm
\includegraphics[width=40pc]{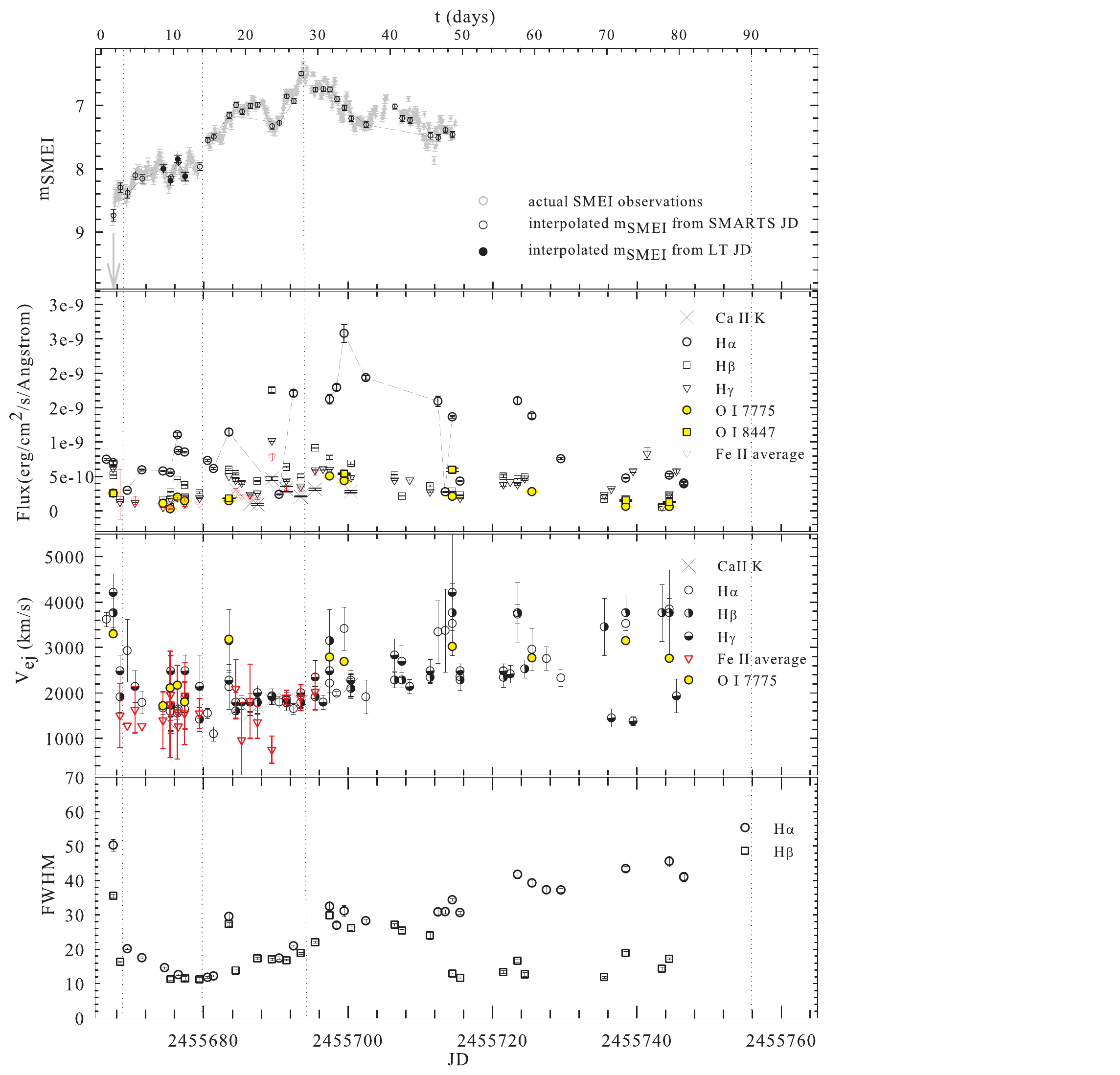} 
\end{center}
\caption[SMEI light curve of T Pyx compared to flux and ejection velocities from Balmer and Fe II lines.]
{SMEI light curve of T Pyx compared to flux and ejection velocity from Balmer, average Fe II (5169, 5018, 4233, 4178, and 4173\AA$ $), O I, and Ca II K lines. Dashed lines in the top two panels connect observations of H$\alpha$ flux. The evolution of the FWHM of Balmer lines is also shown (bottom panel). Vertical dotted lines represent phases in the light curve referred to in the text.}
\label{lc-vej-flux-H-FeII}
\end{figure}

The spectra of T Pyx are also used here to try to understand what causes the gross changes in the light curve - e.g. changes in mass loss rate from the WD surface as the TNR proceeds - and to derive other parameters of the outburst. In order to compare the spectra to the light curve at the exactly the same date, we interpolate the $m_{{\it SMEI}}$ of the date (at the start time of the observations) for which the spectra are available as shown in the top panel of Figure \ref{lc-vej-flux-H-FeII}.

The light curve was compared to the flux and ejection velocities measured from H$\alpha$, H$\beta$, and H$\gamma$ lines as shown in this figure. The variation in flux of the Balmer lines is similar to that found in the light curves before visual maximum. For example, the variation in H$\alpha$ flux around on 2011 May 17 ($t$=33.7 days) is obviously responsible for the sharp and high-amplitude peak light curve at this time.

\subsubsection{Velocities derived}
\begin{figure}[phb!]
\begin{center}
\leavevmode
\epsfxsize = 14.0cm
\epsfysize = 14.0cm
\includegraphics[width=30pc]{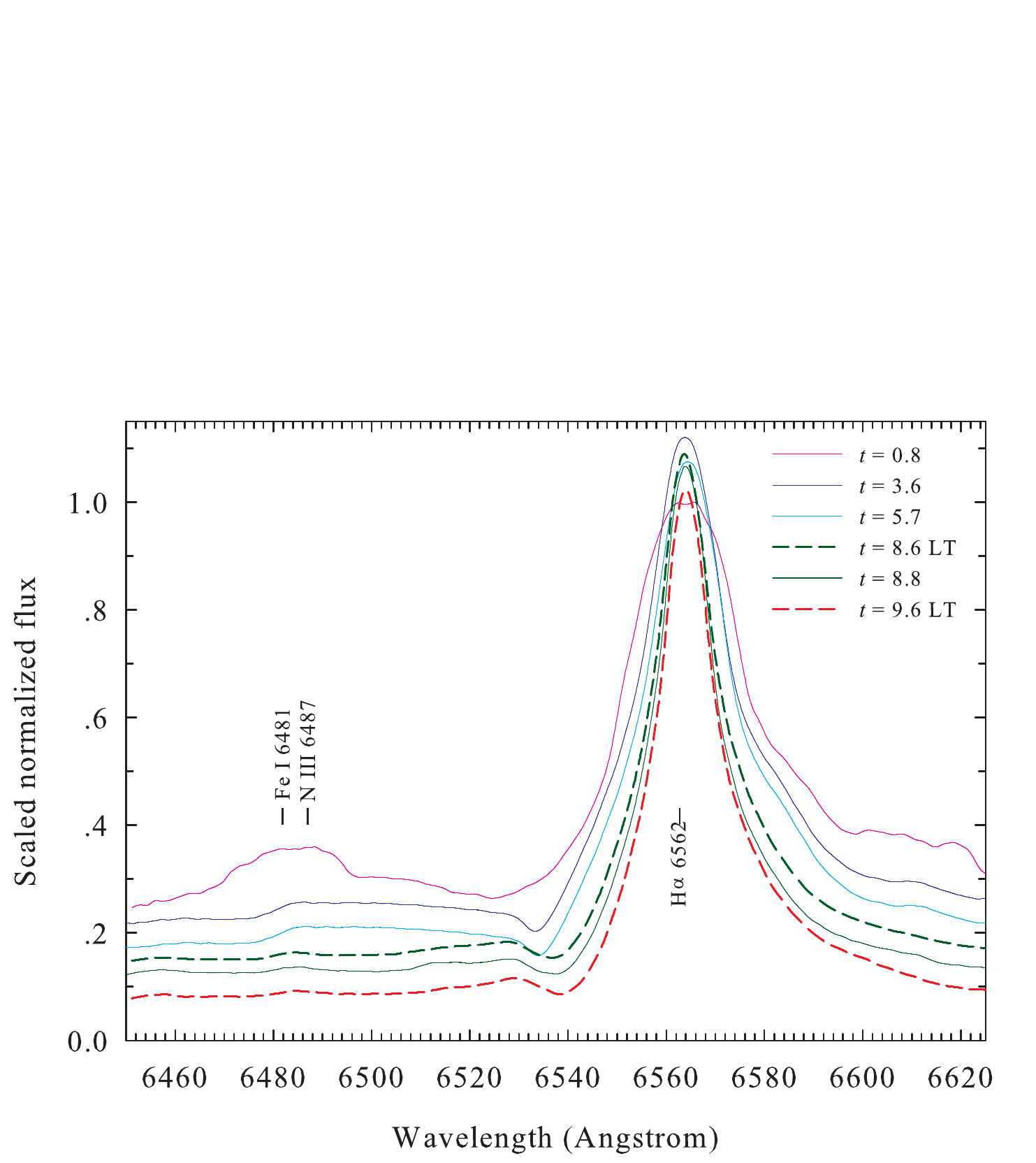} 
\end{center}
\caption[P Cygni profile of H$\alpha$ lines from the initial rise to the end of the pre-maximum halt phase]
{P Cygni profiles of H$\alpha$ lines from the initial rise to the end of the pre-maximum halt phase.}
\label{pcygni-samples}
\end{figure}

The ejection velocities ($V_{ej}$) were measured from the P Cygni profiles. The measured lines include H$\alpha$, H$\beta$, H$\gamma$ which have measurable P Cygni profiles from 2011 Apr 16 - 2011 Jun 1 ($t$=2.7-48.6 days), Fe II recombination lines at 5169, 5018, 4233, 4178, and 4173\AA$ $, O I 7775\AA$ $, and the Ca II K line at 3934\AA$ $. Figure \ref{pcygni-samples} shows the P Cygni profiles of H$\alpha$ lines from the initial rise to the end of the pre-maximum halt phase.

%

\subsubsection{Spectral evolution}

\begin{figure}[phb!]
\begin{center}
\leavevmode
\epsfxsize = 14.0cm
\epsfysize = 14.0cm
\includegraphics[width=28pc]{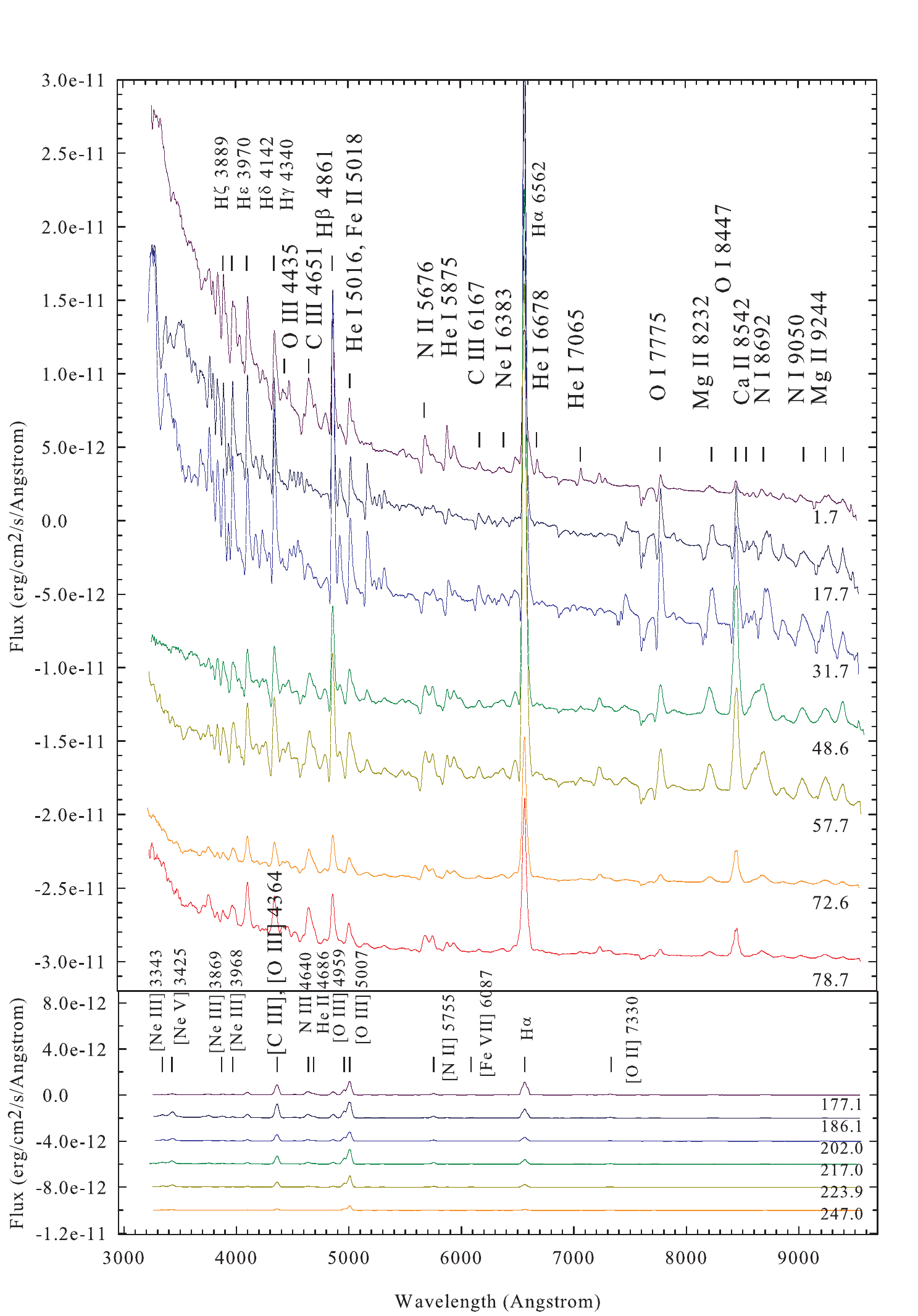} 
\end{center}
\caption[Low-resolution spectra of T Pyx taken from $t$=1.7-247 days.]
{Low-resolution spectra of T Pyx taken from $t$=1.7-247 days. Spectra before the seasonal gap (top) are offset in flux for clarity, as indicated, with the spectrum at 1.7 days representing the observed flux and those from 17.7-78.7 days being offset in steps of 5$\times$10$^{-11}$ erg cm$^{-2}$\AA$^{-1}$. Spectra after the seasonal gap (bottom) are offset in flux with the spectrum at 177.1 days representing the observed flux and those from 186.1-247.0 days being offset in steps of 2$\times$10$^{-11}$ erg cm$^{-2}$\AA$^{-1}$.}
\label{all-low-res}
\end{figure}

Figure \ref{all-low-res} shows all low-resolution spectra taken from $t$=1.7-247 days which cover the initial rise phase through to the transition phase of the light curve. We discuss the spectral evolution based on the idealised nova optical light curve given in \citet{war08} together with a recognition of a common pattern of line development described by \citet{mcl42,mcl44}. We also note the physical interpretation of various stages. For example, from the earliest moment, the nuclear explosion on the surface of the WD leads to the ejection of a hot, luminous, and massive shell that expands radially with time. \citet{ney78} called this stage the $``pseudo-photospheric$ $expansion''$ for the spectral energy distribution (SED) and spectroscopic features are characteristic of the photosphere of a star with spectral type A to F \citep{pay57}. \citet{geh88} called this stage the $``fireball''$ because it has been used to describe the early development of man-made atomic explosions and therefore he used it to describe the expanding pseudo-photosphere of the nova.

    The fireball's envelope is initially small and dense with the radiation peak at X-ray wavelengths. Then the envelope cools adiabatically as increases in size. As a result, the opacity increases, the ejecta become optically thick after the outburst, and the radiation peak shifts towards longer wavelengths \citep{sho94}. Therefore, this cooling of the ejecta envelope together with the rapid expansion of the photosphere \citep{enn77,geh80} provide the rapid initial rise in the optical light curve.

    Following optical maximum, the effective temperature ($T_{eff}$) of a nova, which is radiating at constant bolometric luminosity, rises as the pseudo-photosphere shrinks back onto the WD and as the mass loss rate from the WD decreases \citep{bat89}. As a result, the peak emission now shifts toward shorter wavelengths. The effective temperature of the pseudo-photosphere changes as the visual flux declines according to
    \be
    T_{eff} = T_{0}\cdot10^{\Delta{V}/2.5}  
    \label{Teff}
    \ee
    where $\Delta$$V$ is the decline in magnitude from visual maximum, and $T_{0}$ is the photospheric temperature at optical peak \citep{bat89}. We take $T_{0}$=8000K \citep{eva05} and use the equation above to estimate $T_{eff}$. According to \citet{bat89}, the physical state of the photosphere (i.e. temperature, radius, density, and pressure) is approximately the same at equal magnitudes below the optical maximum in all novae. Consequently, one might anticipate that the spectral appearance of a nova should be the same at the same magnitude below peak in all novae if this simple model is correct. It can then be applied before and after peak if the mass-loss rate from the WD is the governing factor.

\begin{enumerate}
\item {\itshape The initial rise (2011 Apr 14-16, $t$=0.8-3.3 days):\/}
    Accordingly, the spectral evolution is divided into the following four stages. The first red spectra were obtained on 2011 Apr 15.05 UT (JD 2455666.55056, $t$=0.8 days) about 27 days before the visual maximum. At this time, the AAVSO visual magnitude was $V$$\sim$8.7 (about 6.8 mags brighter than quiescence at $V$$\sim$15.5 and about 2.4 mags below visual maximum - see Figure \ref{tpyx-lc-aavso}). The top panel of Figure \ref{r14-b16} reveals that the lines were broad and diffuse. H$\alpha$ and H$\beta$ emissions were strong while the P Cygni profiles were not yet seen clearly when compared to other later spectra, shown in Figure \ref{pcygni-samples}. Moreover, the spectrum at $t$=1.7 days in Figure \ref{all-low-res} shows the brightness of the nova comes almost entirely from the continuum at this early stage of the outburst. This is consistent with the characteristics of the $pre-maximum$ $spectrum$ stage given by \citet{pay57}. This is expected as the expanding pseudo-photosphere is of similar extent to the maximum radius initially reached by the ejecta. 

    The $pre-maximum$ $spectrum$ is defined as the earliest spectrum at which any given nova has been observed on the rise until at least one or two days after maximum light \citep{mcl42}. It usually contains lines which are broad and diffuse with negative velocity displacement \citep{pay57}.

    Our spectra showed P Cygni profiles in the Balmer, Fe II, and O I lines quite early in the first spectra, during this fireball phase. This aspect is similar to that observed at this time in DQ Her and LMC 91 as mentioned in \citet{schw01}. The absorption components of P Cygni profiles became broader and shallower from $t$=0.8 to 2.7 days.

    The initial rise ended just after 2011 Apr 16.95 UT (JD 2455668.44853, $t$=2.7 days) where the first medium-resolution blue spectra were obtained. At $t$=0.8 days, the H I Balmer lines were present in emission and becoming stronger with blue-shifted absorptions and the presence of He I. By $t$=2.7 days, the rise of ionized iron emission lines was evident. There were emissions of high excitation lines (see Table \ref{br-14-23}) i.e. C III, N III, Ne II, O II, N II, He I, and Ne I present during this phase (see Fig \ref{r14-b16}) which disappeared later around 2011 Apr 22-23 ($t$=8.6-9.7 days). These O II and N II lines are also those expected as the ionized elements that should be found at $\sim$1.5 mag below peak \citep{bat89} - see Table \ref{bat89}. Other expected emission lines at 2.9 mag below peak are O III at 4435\AA$ $ which is also seen at $t$=1.7-2.7 days.

\begin{table}
\begin{center}
\caption[Ionization levels at various decline stages according to \citeauthor{bat89} (1989; only species with strong optical lines noted).]
{Ionization levels at various decline stages according to \citeauthor{bat89} (1989, Only species with strong optical lines noted).}
\vspace{0.3cm}
\small
\begin{tabular}{ccccccccccccccccccccccccccccccccccccccccccccccccccc}
\hline \hline
$\Delta$$B$ & $B$& Element ionization level   & $t$ (days)                 & $T_{eff}$ (K)\\
\hline
1.45        & 7.78 & O II                       & 2.7, 3.6-13.6, 33.7-61.7       & 30,000   \\
1.55        & 7.88 & N II                       & 2.7, 3.6-10.7, 36.7-80.8       & 33,000   \\
2.85        & 9.18 & N III                      & 0.8-2.7, 5.7, 8.6, 73.7, 155.1-221.9 & 110,000   \\
3.15        & 9.48 & O III                      & 0.8-1.7, [45.6]$^{1}$-246 & 145,000    \\
4.85        & 11.18 & N V                   &   73.7 (N V 4603\AA$ $)   & 700,000\\
\hline \hline
\label{bat89}
\end{tabular}
\end{center}
\vspace{-0.7cm}
\small
\begin{center}
$^{1}$ Marginally detected at this time.
\end{center}
\end{table}

    A marked drop in the derived expansion velocity ($V_{ej}$) during this phase is also noted (see Figure \ref{lc-vej-flux-H-FeII}). We note that \citet{ima12} and \citet{sho11} also report a decline in derived expansion velocities at early times. The first observations show a derived expansion velocity of $\sim$4000 km s$^{-1}$ at $t$=0.8 days which then drops to $\sim$2000 km s$^{-1}$ at $t$=2.7 days for Balmer lines. This should not be interpreted as a deceleration. This aspect of the pre-maximum spectrum with the dramatic decrease in $V_{ej}$ during the initial rise is also found in the slow nova DQ Her \citep{mcl37} for Balmer lines. Meanwhile the fast nova V603 Aql \citep{wys40} also showed the decrease in $V_{ej}$ for Balmer and metal lines from a few observations during the final rise.

    If the initial ejection is a Hubble flow, then one will a see high $V_{ej}$ initially which declines as one sees into deeper layers. We also note, as an aside, a comparison to the high velocity features (HVFs) found in all Type Ia SNe and believed to be the result of the interaction of initial highest velocity ejecta with a circumstellar envelope \citep{ben05}. The subsequent change in behaviour of the derived $V_{ej}$ during the pre-maximum halt phase suggests two different stages of mass loss: a short-lived phase first occurring immediately after outburst and then followed by a more steadily evolving and higher mass loss phase.



\begin{figure}[phb!]
\begin{center}
\leavevmode
\epsfxsize = 14.0cm
\epsfysize = 14.0cm
\includegraphics[width=25pc]{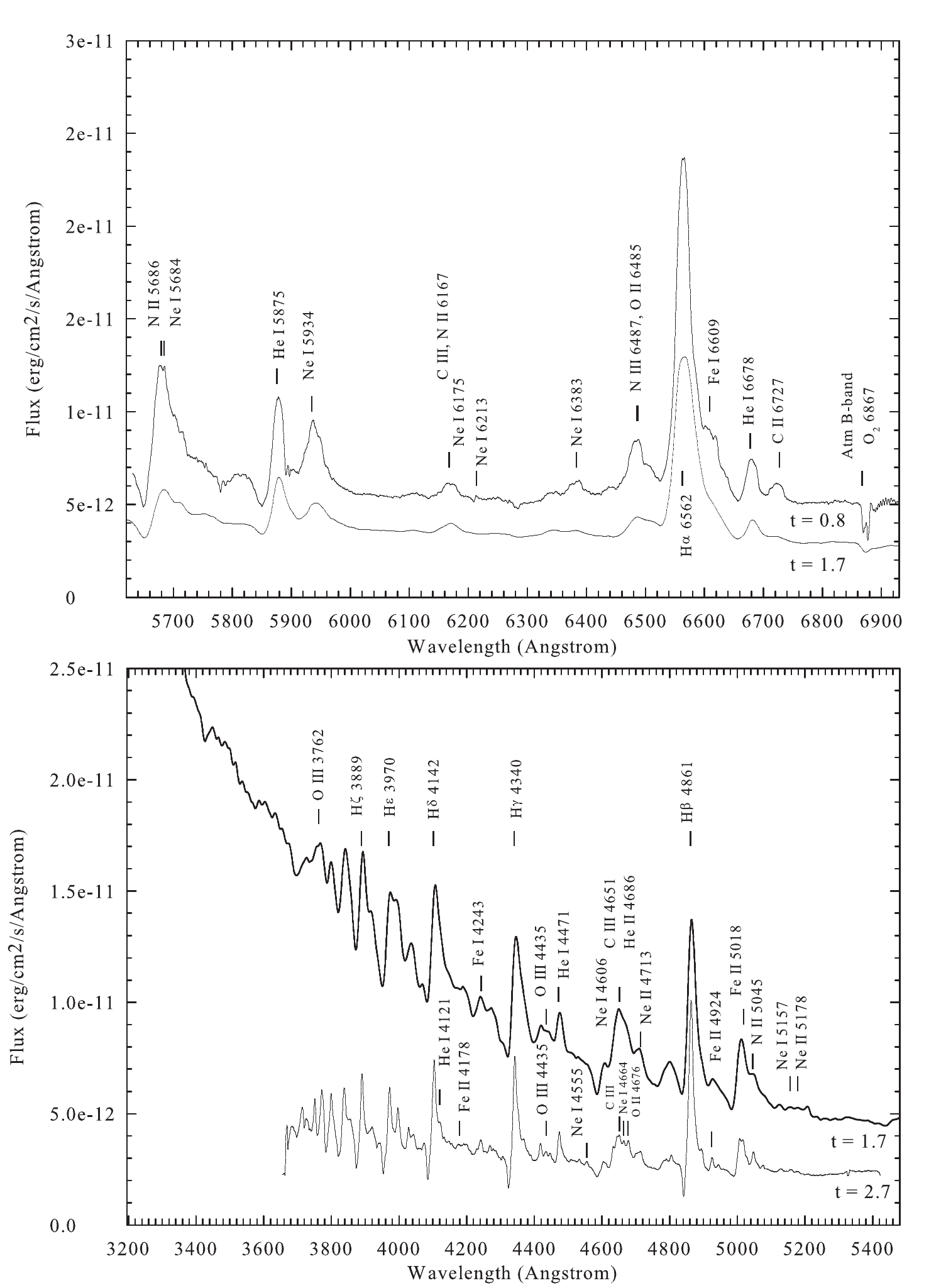} 
\end{center}
\caption[Detection of high ionization lines (e.g. O III and N III) present during the initial rise.]
{Detection of high ionization lines (e.g. O III and N III) present in red spectra (top) and in blue spectra (bottom) during the initial rise ($t$=0.8-2.7 days). Spectra show the observed flux in each case.}
\label{r14-b16}
\end{figure}

\begin{table} [pht!]
\begin{center}
\caption[High ionization lines and some selected lower ionization lines present in the spectra of T Pyx 2011 in its initial rise and pre-maximum halt phases.]
{High ionization lines and some lower ionization lines present in the spectra of T Pyx 2011 in its initial rise and pre-maximum halt phases.}
\vspace{0.3cm}
\small
\begin{tabular}{*5c}
\hline \hline
High excitation lines & Ionization potential (eV) & $\lambda$ (\AA$ $)    & \multicolumn{2}{c}{Days since discovery}  \\
                      &                           &                      & First detected  & Last detected \\
\hline
O III	                    & 54.90                     & 4435.0            & 1.7        & 2.7  \\
C III	                    & 47.90                     & 4651.4            & 1.7        & 3.6  \\
    	                    &                           & 6167.5            & 0.8        & 3.6  \\
			                & 			                    &  6727.0           & 0.8        & 3.6  \\
N III	                    & 47.24                         & 5943.0$\dagger$           & 0.8        & 5.7  \\
     	                    &                               & 6487.0$\dagger$$\dagger$  & 0.8        & 8.6  \\
Ne II				        & 41.00 		                &  4713.4           & 1.7        & 2.7  \\
    				        &            	                &  5178.5$\dagger$$\dagger$$\dagger$ & 1.7 & 2.7  \\
O II	                    & 35.10                     & 4676.2             & 2.7        & 9.7  \\
N II	                    & 29.60                     & 5045.0            & 1.7        & 2.7  \\
			                & 			                    &  5686.2            & 3.6        & 10.7  \\
			                & 			                    &  5938.0            & 3.6        & 11.6 \\
He I	                    & 24.60                     & 4120.8            & 2.7        & 8.6  \\
     	                    &                           & 5875.6            & 0.8            & 14.8  \\
			                & 			                    &  6678.1           & 0.8        & 14.8 or 26.7?  \\
			                & 			                    &  7065.2           & 1.7        & 17.7  \\
Ne I	                    & 21.60                         &  4663.5           & 2.7        & 9.7  \\
			                & 			                    &  5156.6           & 2.7        & 8.6  \\
			                & 			                    &  5684.6           & 0.8        & 3.6  \\
			                & 			                    &  5934.4           & 0.8        & 8.6  \\
			                & 			                    &  6213.8           & 0.8        & 0.8  \\
\hline \hline
\label{br-14-23}
\end{tabular}
\end{center}
\vspace{-0.7cm}
\small
\begin{center}
$\dagger$ N III 5943\AA$ $  + Ne I 5934\AA$ $ could account for N II 5938\AA$ $. \\
$\dagger$$\dagger$ Could also be N II 6487\AA$ $.\\
$\dagger$$\dagger$$\dagger$ Could also be Mg I 5178\AA$ $.\\
\end{center}
\end{table}

\item {\itshape The pre-maximum halt (2011 Apr 17-27, $t$=3.6-13.7 days):\/} The SMEI light curve at this phase has $m_{{\it SMEI}}$$\sim$8 mag (with 0.1-0.3 mag variations) which is about 2 magnitudes below maximum (6.33 mag). This phase lasts for about ten days. As \citet{war08} notes, the pre-maximum phase is generally much longer-lasting in slow novae than in fast novae.

    All of the C III lines disappeared by $t$=3.6 days while other lower ionization lines such as N III, Ne II, N II and Ne I lasted a little longer until disappearing approximately around $t$=5.7-10.7 days as shown in Table \ref{br-14-23}, consistent with behaviour reported in \citet{sho11}. The spectra at $t$=3.6-5.7 days show broad and strong emission lines present between $\lambda$$\sim$6480-6530\AA$ $ and then these fade away. Lines which were present all the time during this phase are  O I, O II, He I and Fe I which are again consistent with the behaviour proposed by \citet{bat89} at this magnitude below peak. We note that He I 5875\AA$ $ and 6678\AA$ $ were present from $t$=0.8 days then became weaker and totally disappear by $t$=14.8 days. They appeared again at $t$=80.7 days.

    The end of the $fireball$ stage occurs around $t$=8.6 days which is also the beginning of the next optically thick phase called the ``$iron$ $curtain$''. The iron curtain \citep{hau92,schw01} occurs when the pseudo-photosphere reaches its minimum temperature ($\sim$10$^{4}$K). The iron-peak elements recombine at this temperature. The overlapping of absorptions of Fe II lines in the near UV region results in line blanketing in the UV which dominates the UV SED and redistributes most of the emitted light into the optical and IR.

    At this stage, we found three characteristics of the iron curtain which are mentioned in \citet{schw01}. First, is the increase in the width of the emission lines which is clearly occuring in the final rise (see Figure \ref{lc-vej-flux-H-FeII}); second, is the increase in the derived expansion velocity which \citet{schw01} suggested could be due to the gradually accelerating optically thick wind that is proposed to begin after the initial outburst \citep{kov98}; finally, the increase in the intensity of Fe II multiplets and O I 7775\AA$ $ and 8447\AA$ $. Figure \ref{b-plateau} shows that the low excitation lines such as Fe-peak transitions, particularly Fe II recombination lines at 5169, 5018, 4233, 4178, 4173\AA$ $, begin to rise from $t$=8.6 days. 

    There is also a tendency for the absorption components, especially those of the Fe II lines, to become sharper and stronger as maximum is approached. We note that during this phase (for particular ions) the spectrum tends to develop such that the shorter wavelength lines are evident first and then followed by the next lines toward the longer wavelength. For example, Fe I 7443\AA$ $ develops first, followed by Fe I 7446\AA$ $. This also happens with Fe I 7469-7473\AA$ $, O II 7895-7898\AA$ $, and Fe I 8468-8471\AA$ $.

    The variation of the light curve during the pre-maximum halt phase seems to be consistent with variations in the strength of the H$\alpha$, H$\beta$, H$\gamma$, and Fe II lines during this halt as shown in Figure \ref{lc-vej-flux-H-FeII}. There is an obvious sharp variation in the light curve during $t$=8.6-11.6 days where spectra show some Fe I lines were present at $t$=9.6-10.6 days but not before or after that (bottom panel of Figure \ref{r-plateau}). For example, the Fe I 7854\AA$ $ line was not detected at $t$=8.6 days but was present on days 9.6 and 10.6 before disappearing again at $t$=11.6 days. Meanwhile Fe I 8838\AA$ $ was present only at $t$=10.6 days.

    \citet{ima12} and \citet{ede13} suggested that T Pyx evolved from a He/N to an Fe II-type nova \citep{wil92} by the time it reached visual maximum. Here we can point out the exact time that this process began was during this pre-maximum halt phase and was completed by the final rise. We note, however, that this spectral classification is normally applied to novae from maximum light onwards. The evolution of T Pyx from a He/N nova to an Fe II type nova suggests that there was sufficient mass loss for the optical thickness in the outer layers to increase during the rise to maximum so that we were seeing less deep into the expanding layers. This type of evolution could be typical for many novae in outburst but has not been seen before because of insufficient observations early in the outburst.

    The maximum $V_{ej}$ of approximately 2200 km s$^{-1}$ during this phase derived from the Balmer lines agrees very well with that obtained by \citet{ima12}. Although we do not have information from the P Cyg profiles in Fe II lines until $t$=3.6 days to verify the very high $V_{ej}$ at the same initial rise phase as the Balmer lines, the Fe II lines seem to show a similar trend subsequently to that of the Balmer lines in that $V_{ej}$ stabilises at $\sim$1500 km s$^{-1}$ during the pre-maximum halt and tends to increase afterward.

\begin{figure}[h!]
\begin{center}
\leavevmode
\epsfxsize = 14.0cm
\epsfysize = 14.0cm
\includegraphics[width=30pc]{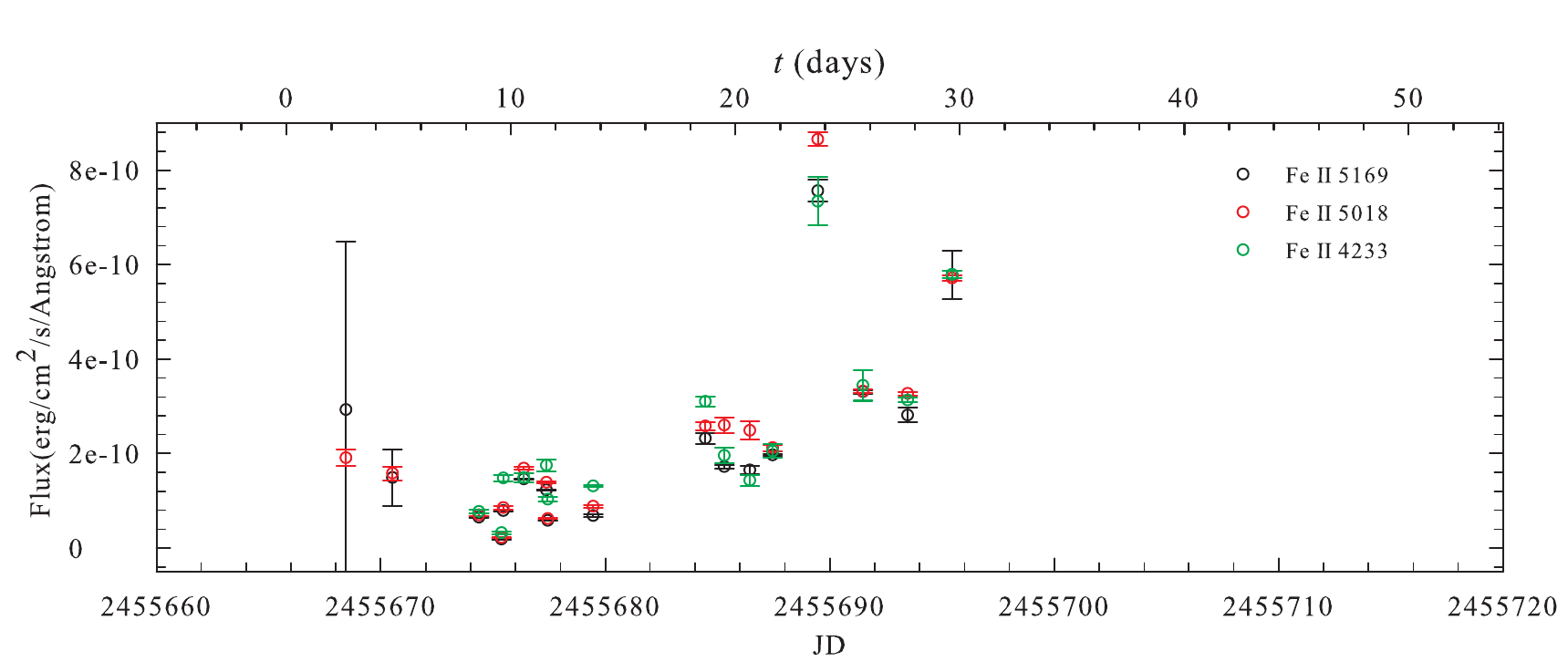} 
\end{center}
\caption[Evolution of the flux of Fe II lines at 5169, 5018, and 4233\AA$ $.]
{Evolution of the flux of Fe II lines at 5169, 5018, and 4233\AA$ $.}
\label{FeII-flux-iron-curtain}
\end{figure}

\begin{figure}[phb!]
\begin{center}
\leavevmode
\epsfxsize = 14.0cm
\epsfysize = 14.0cm
\includegraphics[width=30pc]{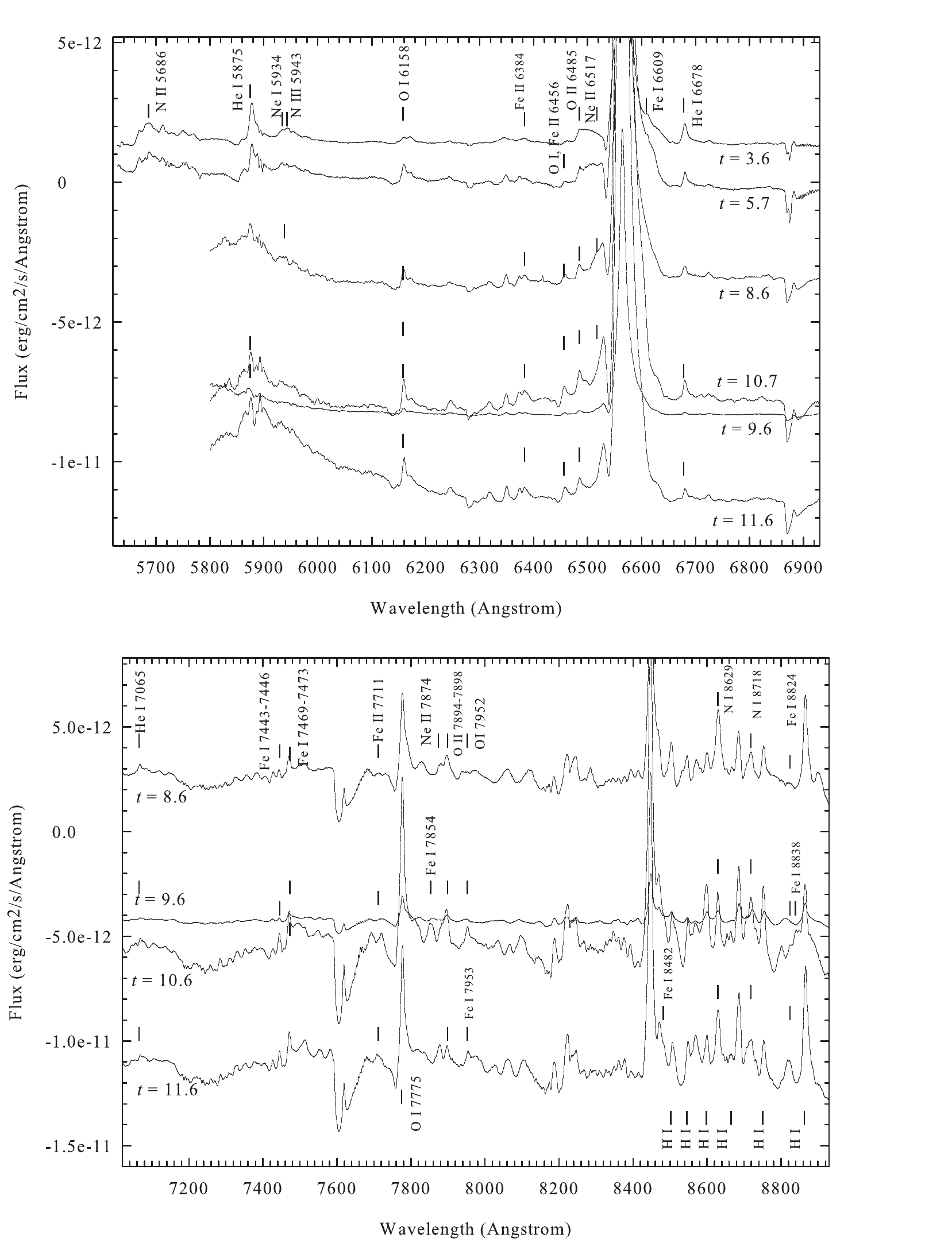} 
\end{center}
\caption[The red spectra during the pre-maximum halt phase.]
{The red spectra during the pre-maximum halt phase. Spectra are offset for clarity in flux with the spectrum at 3.6 days (top panel) and 8.6 days (bottom panel) representing the observed flux and the later spectra being offset in steps of 3$\times$10$^{-12}$ and 5$\times$10$^{-12}$ erg cm$^{-2}$\AA$^{-1}$ in each figure, respectively.}
\label{r-plateau}
\end{figure}

\begin{figure}[pht!]
\begin{center}
\leavevmode
\epsfxsize = 14.0cm
\epsfysize = 14.0cm
\includegraphics[width=30pc]{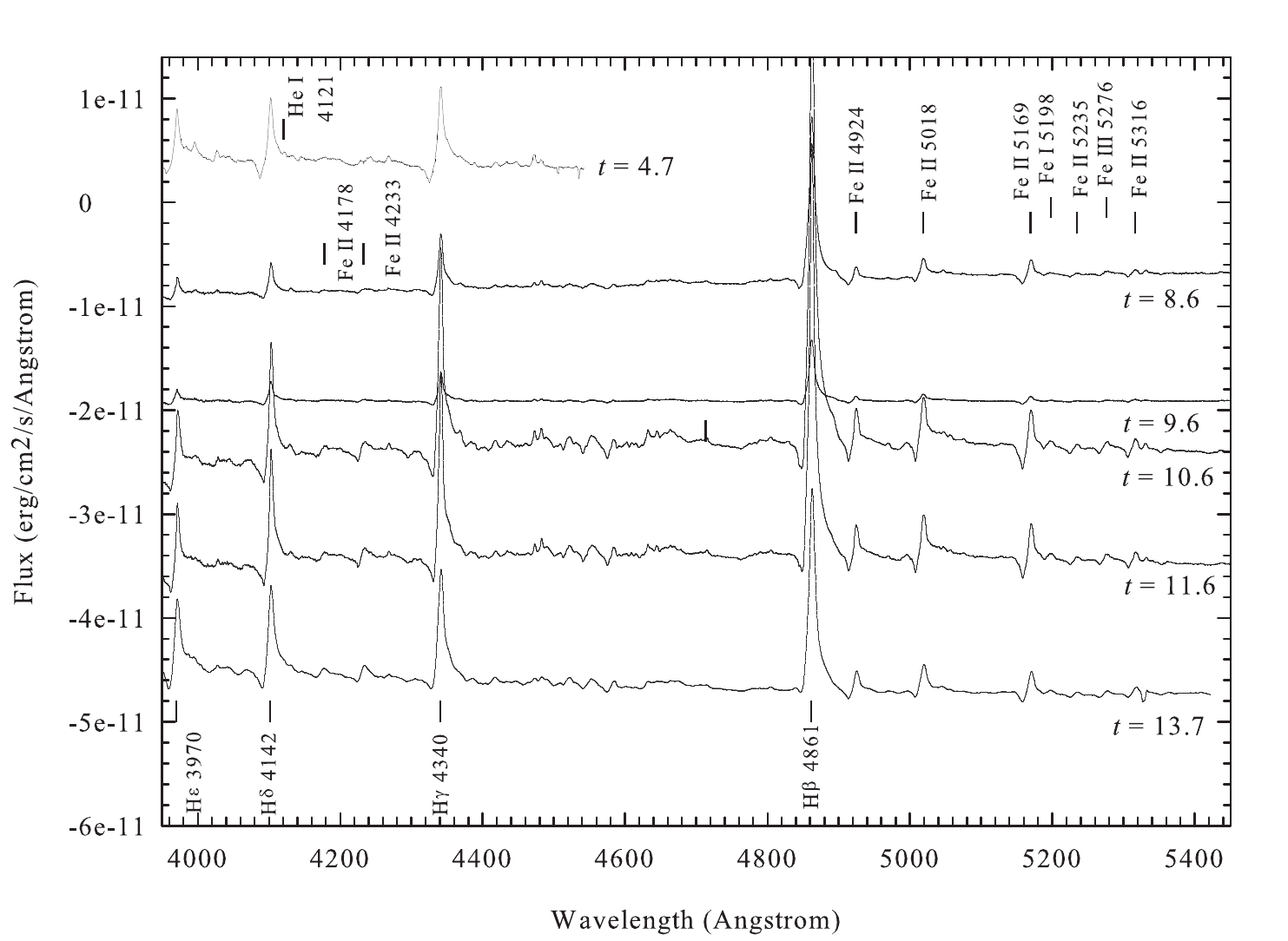} 
\end{center}
\caption[The blue spectra during the pre-maximum halt phase.]
{The blue spectra during the pre-maximum halt phase. Spectra are offset in flux for clarity with the spectrum at 4.7 days representing the observed flux and the later spectra being offset in steps of 1$\times$10$^{-11}$ erg cm$^{-2}$\AA$^{-1}$.}
\label{b-plateau}
\end{figure}

\item {\itshape Final rise (2011 Apr 28 - May 11, $t$=14.7-27.9 days):\/} Here the light curve rises more steeply toward maximum. As noted above, the iron curtain stage is expected to peak around this time. However Figure \ref{FeII-flux-iron-curtain} shows the Fe II flux rose to a sharp peak just before $t$$\sim$23 days when there is a major dip in the SMEI light curve with a minimum at $t$=23.7 days. The visual maximum then occurred four days later at $t$=27.9 days.

    The ``$principal$ $spectrum$'' which dominates CNe spectra at visual maximum \citep{pay57,war08} displays strong lines of O I (i.e. the ``O I flash''). In T Pyx these became apparent around $t$$\sim$17 days and grew in intensity at about the same rate as the Fe II lines. At the O I flash, the $V$ magnitude was $\sim$1.5 mag below peak which agrees with that expected by \citet{bat89} from the $\Delta$$B$ at this time. Strong bright lines of Fe II and Ca II are always present. The emission of [N II] 5755\AA$ $ begins to grow stronger at $t$=26.7 days, about ten days after the O I flash , since N lines, i.e. [N II] 5755\AA$ $ together with emission lines of N III, N IV and N V in the UV, are expected to be seen in the principal spectrum phase of novae \citep{jas09}\footnote{The section (about emission lines of NII in novae) published in the book is available at http://ned.ipac.caltech.edu/level5/Glossary/Jaschek/N.html}.

    Again the Balmer lines and other lines show a similar trend of $V_{ej}$, i.e. gradually increasing (after the initial decrease at early times). This may imply that here the innermost layers of material now move faster than the outer layers (i.e. the relative radius of the pseudo-photosphere to that of the ejecta was shrinking significantly and therefore revealing higher velocity material again). As noted above, this was proposed by \citet{schw01} in terms of a gradually accelerating wind in nova LMC 1991 and has been used to model early hard X-ray emission in some CNe (e.g. \citeauthor{obr94}, 1994). The visual maximum at $t$=27.9 days seems to exhibit the lowest ionization lines, again as predicted in the simple \citet{bat89} models.

    The three characteristics of the iron curtain which are first mentioned in the pre-maximum halt phase persist in this final rise phase. Returning to the major dip in the light curve at $t$=23.7 days, we find that the normalized flux of the blue spectra with respect to the H$\alpha$ line on the day before, at, and after the dip (at $t$=21.7, 23.7, and 25.7 days respectively) all looked exactly the same in shape but only difference in the H$\alpha$ flux that dropped very significantly at the dip. Thus the mechanism that causes the very marked drop in emission line strength of H$\alpha$, and other lines at longer wavelengths, must be responsible for the major dip in the total flux in the final rise. Meanwhile, there are almost no other emission lines apart from H$\alpha$ in the red spectra at $t$=24.7 days.

\begin{figure}[h!]
\begin{center}
\leavevmode
\epsfxsize = 14.0cm
\epsfysize = 14.0cm
\includegraphics[width=30pc]{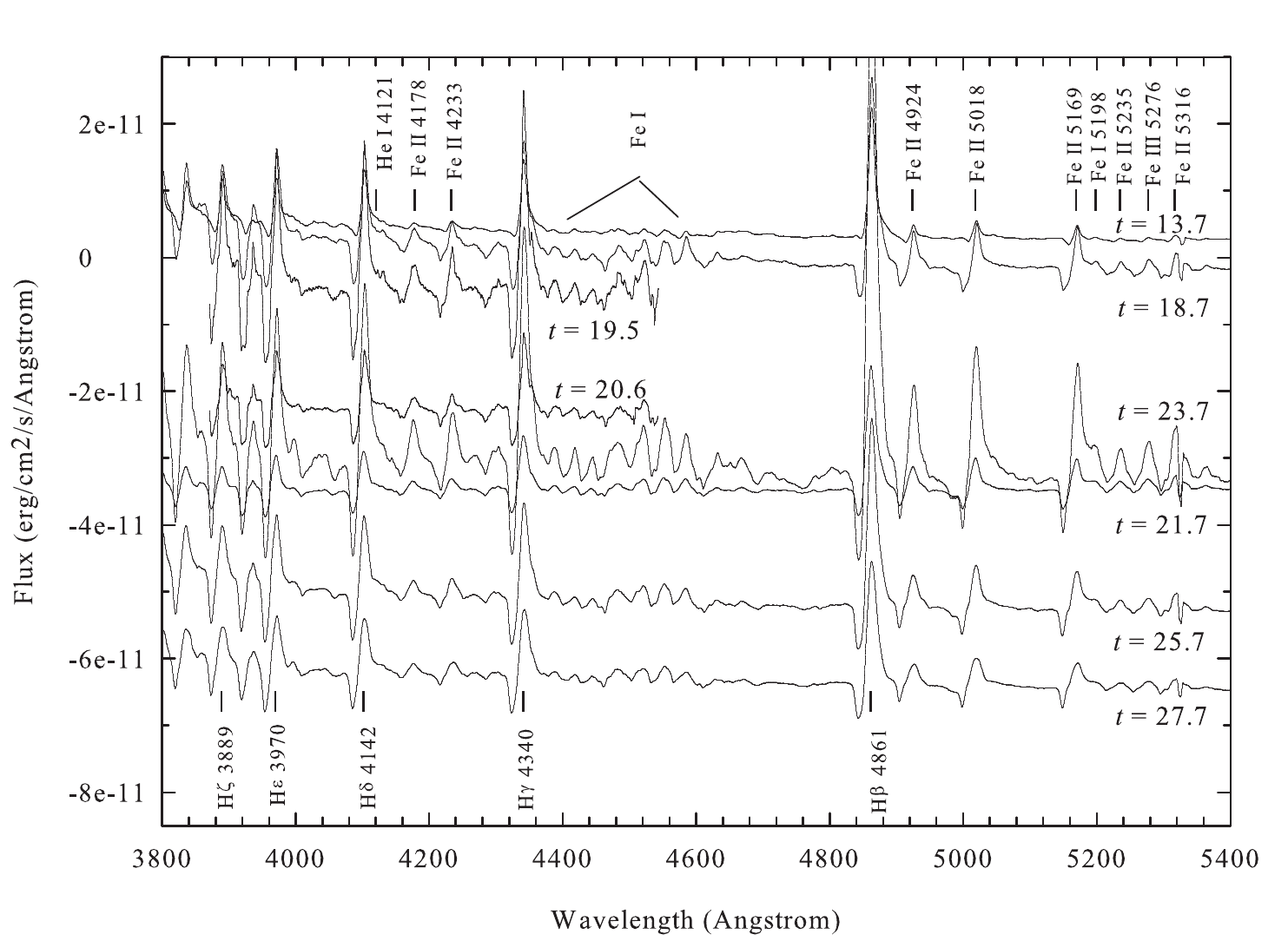} 
\end{center}
\caption[The blue spectra during the final rise phase.]
{The blue spectra during the final rise phase. Spectra are offset in flux for clarity with the spectrum at 13.7 days representing the observed flux and the later spectra being offset in steps of 1$\times$10$^{-11}$ erg cm$^{-2}$\AA$^{-1}$.}
\label{b-rise}
\end{figure}

\begin{figure}[h!]
\begin{center}
\leavevmode
\epsfxsize = 14.0cm
\epsfysize = 14.0cm
\includegraphics[width=30pc]{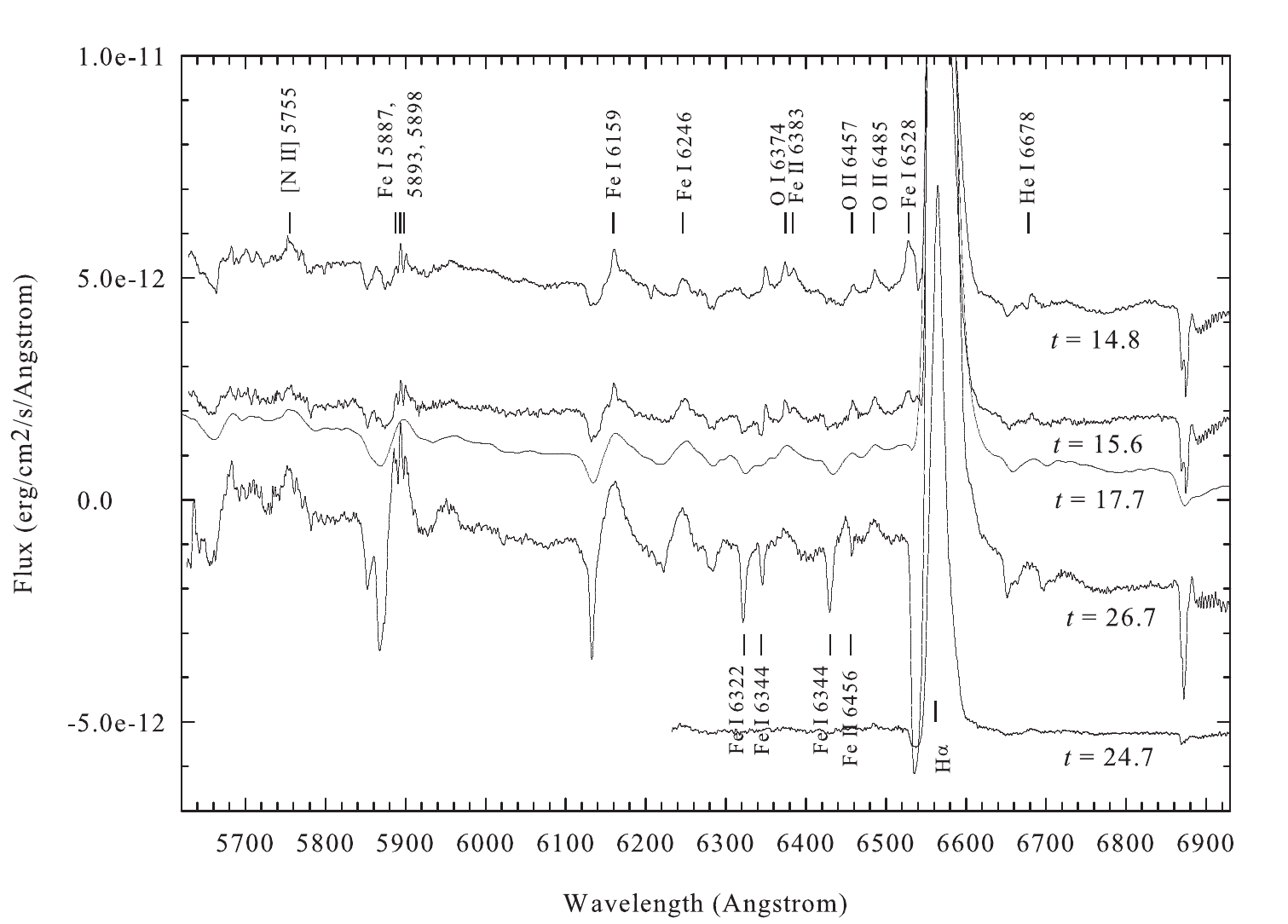} 
\end{center}
\caption[The red spectra during the final rise phase.]
{The red spectra during the final rise phase. Spectra are offset in flux for clarity with the spectrum at 14.8 days representing the observed flux and the later spectra being offset in steps of 2$\times$10$^{-12}$ erg cm$^{-2}$\AA$^{-1}$.}
\label{r-rise}
\end{figure}

\item {\itshape Early decline (2011 May 11 - Oct 3, $t$=27.9-90 days):\/}

    The early decline phase of the CN light curve is defined as the beginning of the decrease in brightness to $\sim$3.5 magnitudes below peak \citep{war08}. The SMEI light curve declines in brightness from $m_{{\it SMEI}}$=6.5 ($t$=27.9 days) to $m_{{\it SMEI}}$=7.26 at the last SMEI detection ($t$=49.7 days). The multi-colour light curves from AAVSO subsequently show sharp drops in all colours at $t$$\sim$90 days, where $\Delta$$V$$\sim$2.5-3.5 magnitudes from maximum.

    H$\alpha$ reached its maximum observed flux at $t$=31.7 days, which is after visual maximum was reached at $t$=27.9 days. The strong emission of Fe II and Ca II, which is always present in the conventional principal spectrum \citep{pay57}, persist until sometime before $t$$\sim$48 days. About five days after optical maximum, the nebular lines of [O I] 6300 and 6363\AA$ $ appeared. The O II lines, which are the expected to became apparent at $\Delta$$B$$\sim$1.5 mag below peak \citep{bat89}, are always present from $t$=33.7-61.7 days.

    The Balmer lines began to show a double-peaked structure from the spectrum taken at $t$=42.7 days. About twenty days after the maximum ($t$=45.6 days), the forbidden line of [O III] 5007\AA$ $ appeared. The flux of Balmer lines began to fade around $t$=70 days. The ``4640 emission'' \citep{pay57} produced by a blend of NIII and NII lines and known as the characteristic of the ``$Orion$ $spectrum$'' becomes apparent at $t$$\sim$70-80 days as shown in Figure \ref{b-decline}. The emission line of N V at 4603\AA$ $, which is expected to appear during the Orion spectrum phase of a typical nova \citep{jas09}, also begins to emerge at $t$=73.7 days. The emergence of these lines is designated the `nitrogen flaring'.  Thus we find that the $Orion$ $spectrum$ stage \citep{pay57,war08} started around $t$=70 days. We note that the N III and N V lines appeared earlier than would have been expected from the simple \citeauthor{bat89} model.

    This evolution is in line with the progression toward the ``$nebular$ $spectrum$'' stage of Classical Novae \citep{war08}. Meanwhile the pseudo-photosphere is continuing to shrink in radius and the effective temperature is increasing.

    The He I 5876\AA$ $ line and the [Fe X] 6375\AA$ $ coronal line were marginally detected at $t$=80.7 days which was the last spectrum observed before the seasonal gap. When spectroscopic observations started again at $t$=155.1 days the nova displayed the expected nebular spectrum.

\begin{figure}[pht!]
\begin{center}
\leavevmode
\epsfxsize = 14.0cm
\epsfysize = 14.0cm
\includegraphics[width=30pc]{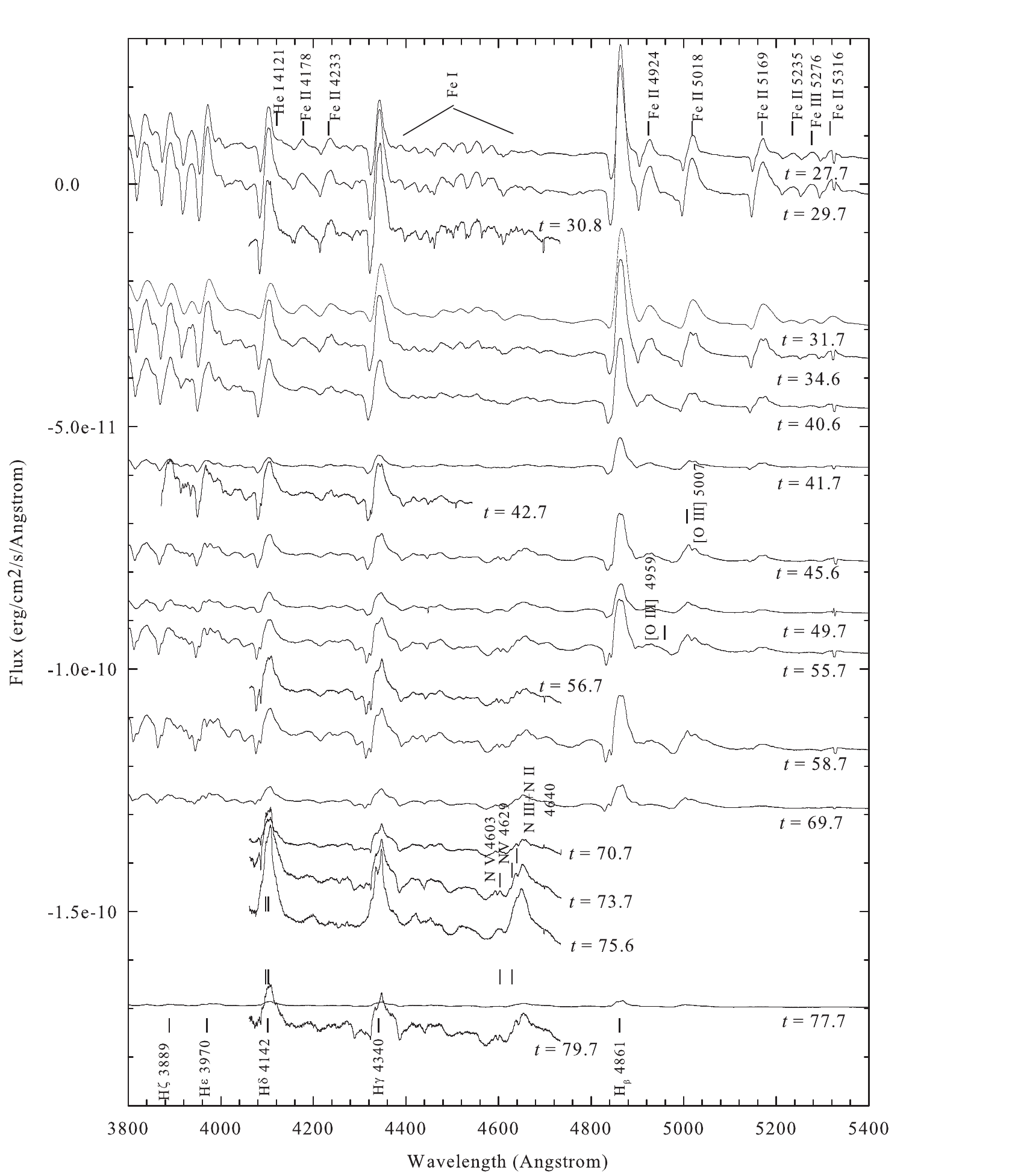} 
\end{center}
\caption[The blue spectra during the early decline phase.]
{The blue spectra during the early decline phase. Spectra are offset in flux for clarity with the spectrum at 27.7 days representing the observed flux and the later spectra being offset in steps of 1$\times$10$^{-11}$ erg cm$^{-2}$\AA$^{-1}$.}
\label{b-decline}
\end{figure}

\begin{figure}[pht!]
\begin{center}
\leavevmode
\epsfxsize = 14.0cm
\epsfysize = 14.0cm
\includegraphics[width=30pc]{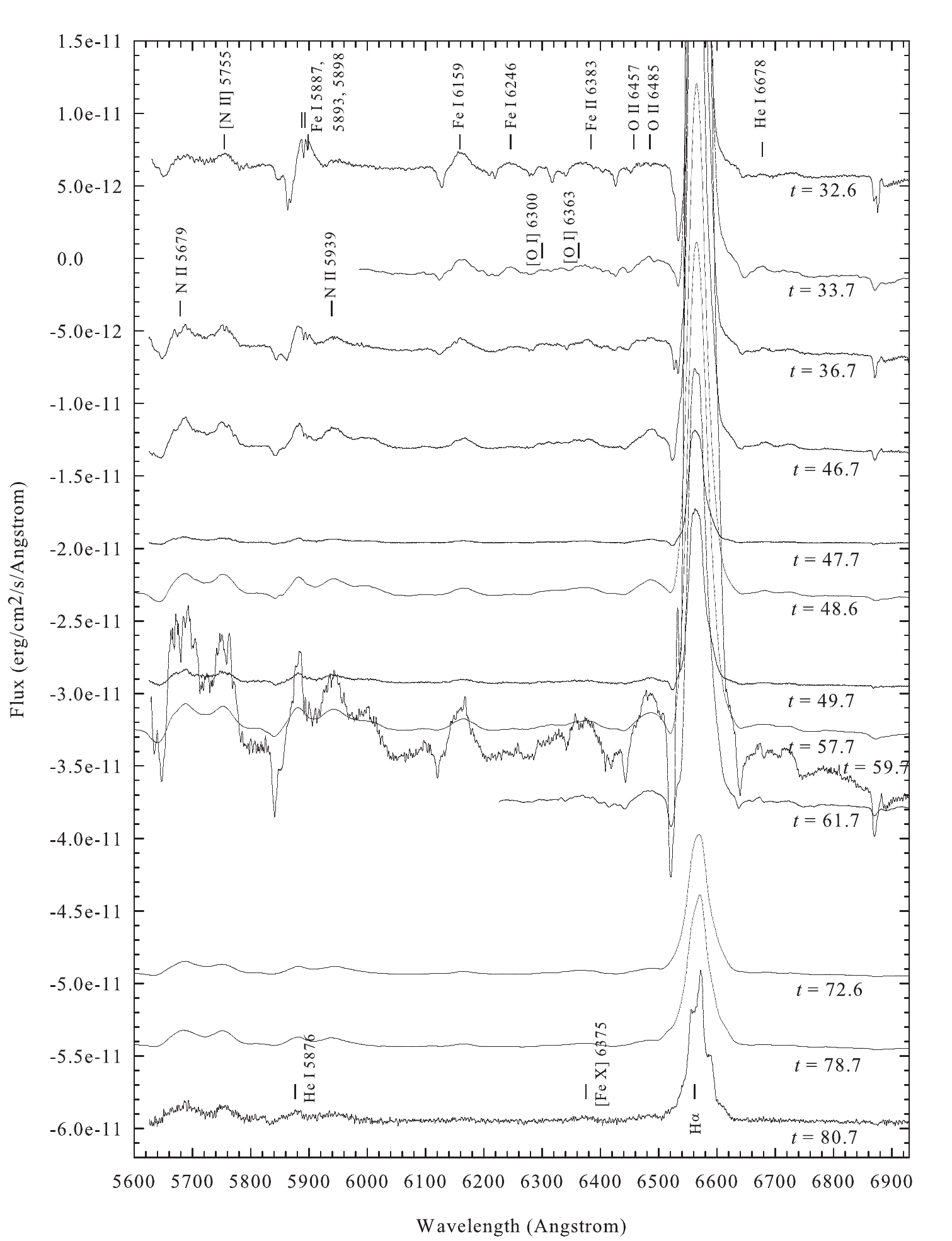} 
\end{center}
\caption[The red spectra during the early decline phase.]
{The red spectra during the early decline phase. Spectra are offset in flux for clarity with the spectrum at 32.6 days representing the observed flux and the later spectra being offset in steps of 1$\times$10$^{-11}$ erg cm$^{-2}$\AA$^{-1}$.}
\label{r-decline}
\end{figure}

\item {\itshape Transition to the nebular phase (2011 Oct 3 - Dec 19, $t$=90-250 days):\/} After the seasonal gap, T Pyx had declined to about 5 magnitudes below maximum. By this time, as indicated in the previous phase, the [O III] 5007\AA$ $ nebular and [Fe X] 6375\AA$ $ coronal lines had clearly developed. It is then of interest to note that the Swift satellite detected the rise in the X-ray light curve at $t$=111 days (see below).



    The first blue spectrum ($t$=155.1 days) already showed [Ne III] 3869\AA$ $, [C III] 4364\AA$ $, N III 4640\AA$ $  and these were increasing in intensity, especially the [C III] line as shown in Figure \ref{b-transition}.  The appearance of N III at $t$$\sim$90 days is consistent with the expectations of \citet{bat89} for $\Delta$$B$=2.9 mag. H$\epsilon$, H$\delta$, and H$\gamma$ have already faded. Moreover, the coronal lines [Fe X] 6375\AA$ $ and [Fe VII] 6087\AA$ $ were clearly present in the first red spectrum at $t$=161.1 days as shown in Figure \ref{r-transition}. The Balmer lines are still stronger than [O III] 5007\AA$ $, however, and they present a multi-peak structure. [O III] 5007 was the strongest rival to the Balmer lines at $t$=165.1 days. The emission of [O III] and N III was strongest again at $t$=221.9 days. The last spectroscopic observation was made at $t$=249.9 days where the nova was clearly in the nebular spectrum stage and still exhibited [NII], He I, [Fe VII], and [Fe X] lines.

\begin{figure}[pht!]
\begin{center}
\leavevmode
\epsfxsize = 14.0cm
\epsfysize = 14.0cm
\includegraphics[width=30pc]{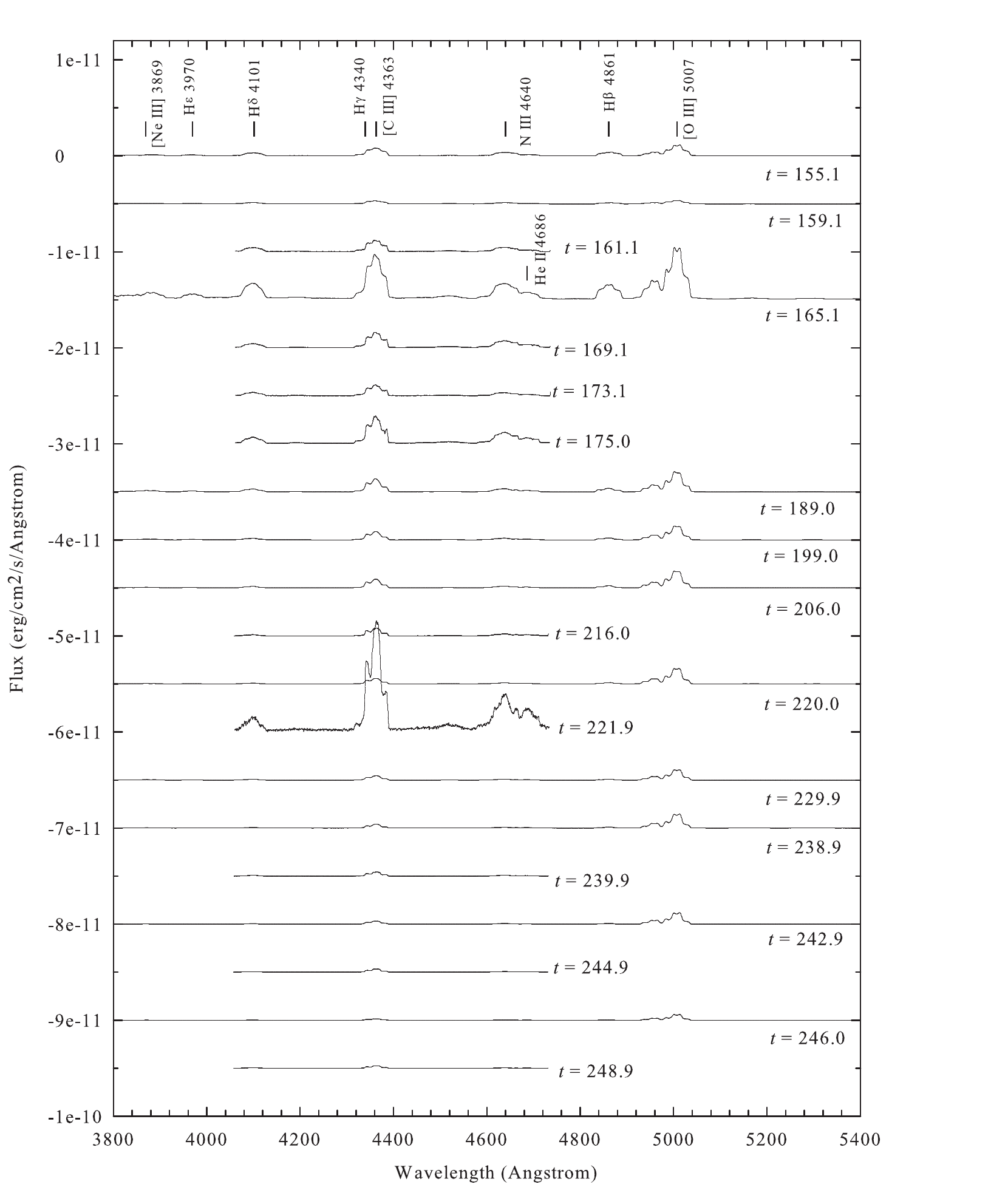} 
\end{center}
\caption[The blue spectra during the transition phase.]
{The blue spectra during the transition phase. Spectra are offset in flux for clarity with the spectrum at 155.1 days representing the observed flux and the later spectra being offset in steps of 5$\times$10$^{-12}$ erg cm$^{-2}$\AA$^{-1}$.}
\label{b-transition}
\end{figure}

\begin{figure}[pht!]
\begin{center}
\leavevmode
\epsfxsize = 14.0cm
\epsfysize = 14.0cm
\includegraphics[width=30pc]{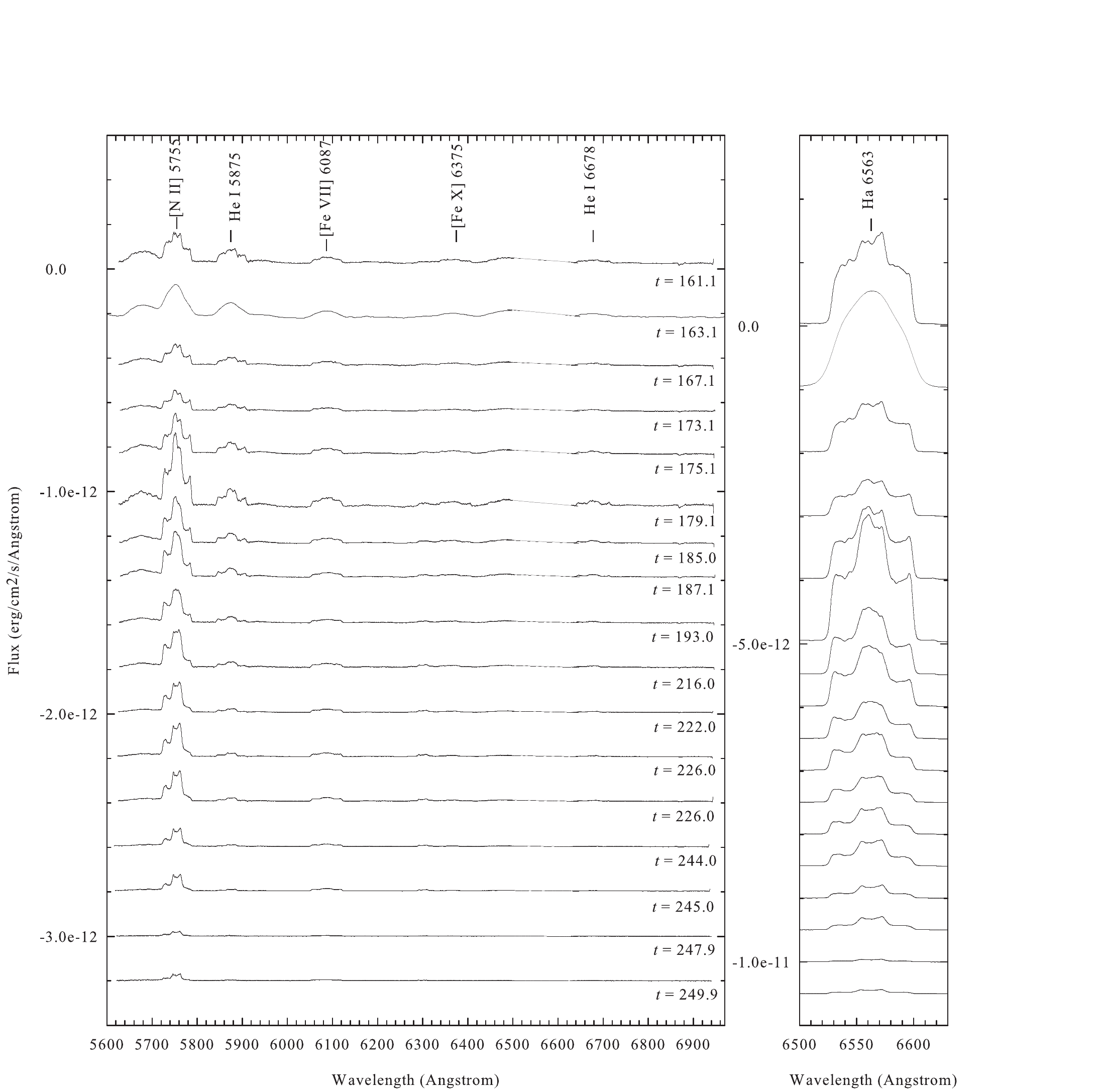} 
\end{center}
\caption[The red spectra during the transition phase.]
{The red spectra during the transition phase. Spectra are offset in flux for clairity with the spectrum at 161.1 days representing the observed flux and the later spectra being offset in steps of 2$\times$10$^{-13}$ erg cm$^{-2}$\AA$^{-1}$ for the left panel and steps of 1$\times$10$^{-12}$ erg cm$^{-2}$\AA$^{-1}$ for the right panel.}
\label{r-transition}
\end{figure}

\end{enumerate}

\newpage
\begin{sidewaysfigure}
\begin{center}
\leavevmode
\epsfxsize = 14.0cm
\epsfysize = 14.0cm
\includegraphics[width=55pc]{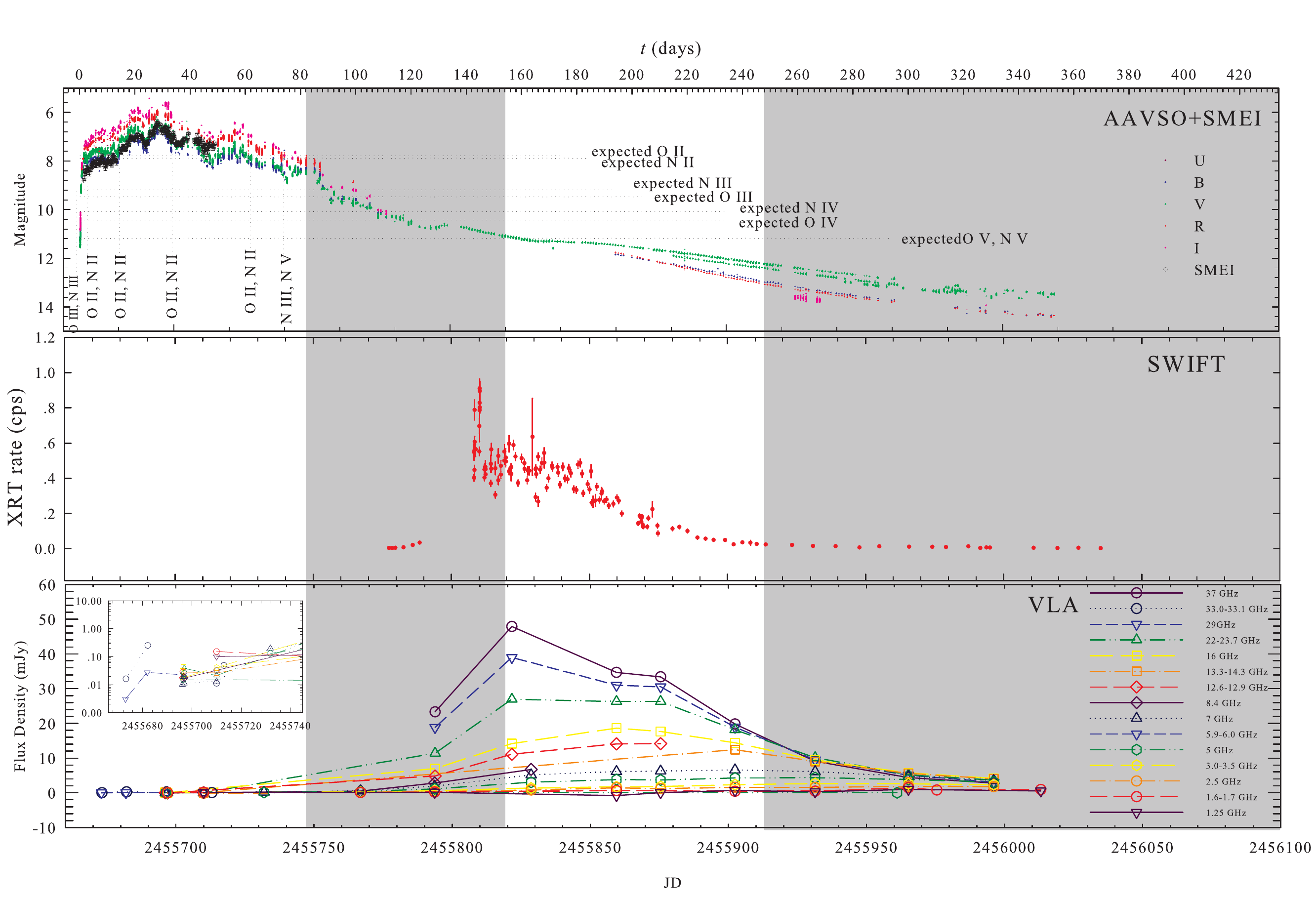} 
\end{center}
\caption[Comparison between optical, radio and X-ray light curves.]
{Comparison between optical, radio and X-ray light curves with the top panel also indicating the expected appearance of various ionized species from the models of \citet{bat89}. Shaded areas represent the epochs where we do not have spectroscopic coverage. The early evolution of the radio flux is shown in the inset in the bottom panel with a logarithmic flux scale for clarity.}
\label{radio-x-lc}
\end{sidewaysfigure}

\section{Comparison to the X-ray and Radio Light Curves}
Figure \ref{radio-x-lc} shows the optical light curves from AAVSO and SMEI, together with the expected appearance of various spectral lines from \citet{bat89}, the X-ray light curve from SWIFT, and the radio light curves from the VLA as presented in \citet{nel12}. During the pre-maximum halt phase in the SMEI light curve, the rise of the high frequency (33 GHz) radio emission was detected at $t$=7-15 days but X-rays were not yet detected. During the final rise through the early decline (until $t$$\sim$45 days) radio emission at all frequencies tended to be stable with a small trend of increase during the optical decline. At $t$$\sim$45 days, [O III] 5007\AA$ $ appeared, and while the radio emission subsequently rose steeply, the X-rays were still not detected.

The Swift satellite detected the rise of X-ray emission (0.3-10 keV) at $t$=111 days, at $\Delta$$V$$\sim$ 4 mag below peak. From Equation \ref{Teff}, $T_{eff}$ would be $\sim$320,000 K. This is typical for a SSS (see \citeauthor{kah06}, 2006). Taking $L_{bol}$$\sim$2$\times$10$^{38}$ erg s$^{-1}$, calculated from M$_{bol}$=$-$7.0 given by \citet{sch10b}, the approximate radius of the pseudo-photosphere at this time would be 5.1$\times$10$^{9}$cm. As expected, this is smaller than the binary separation calculated from the binary parameters given in \citet{uth10} as 6.1$\times$10$^{10}$cm, but larger than the radius of $\sim$4.9$\times$10$^{8}$cm of a 1$M_{\odot}$ WD \citep{sta12}. We note that although the Chandra grating spectra at this phase showed that emission lines were very strong in the X-ray spectrum \citep{ori12}, the SSS continuum was detected in the grating spectra at $t$=210 days by \citet{tof13}, consistent with emission from a WD using a model atmosphere temperature $\sim$420,000 K. The X-ray emission at this time appears to be a mix of a SSS and shocked circumstellar gas (probably from intra-ejecta shocks - \citealp{tof13}).

The X-ray emission then rose to a peak at $t$$\sim$144 days \citep{kuu11}. This is consistent with the appearance of lines from highly ionized species such as [Ne III] 3869\AA$ $, [C III] 4364\AA$ $, and N III 4640\AA$ $, that we found in the first blue spectrum ($t$=155.1 days) right after the seasonal gap. The presence of the N III line was expected some time between 85 and 90 days in the seasonal gap according to $\Delta$$B$ \citep{bat89} although N III+ N II 4640\AA$ $ was detected slightly earlier at 73.7 days. By the time of the rise in X-rays H$\epsilon$, H$\delta$, and H$\gamma$ had already faded.

The clear presence of the coronal lines [Fe VII] 6087\AA$ $ and [Fe X] 6375\AA$ $ at $t$=161.1 days coincides with the middle of the X-ray plateau phase and also the peak of the radio emission (see below). The X-ray emission observed by Swift again became undetectable at $t$$\sim$222 days but Chandra observations still detected the composite line and continuum X-ray spectrum on $t$=235 days, although at least 50$\%$ lower in flux than on $t$=210 days \citep{tof13}. The blue optical spectrum at $t$=221.9 days shows the strongest emission to be [C III] 4364\AA$ $ and N III 4640\AA$ $.

The radio light curve kept increasing throughout the X-ray rise and peaked at around $t$$\sim$155 days for the highest frequency (37 GHz) corresponding to the middle of the plateau in the X-ray light curve. In contrast, the lowest frequency (1.25 GHz) seemed to reach its peak at $t$=290-330 days. \citet{nel12} suggested that it was the material ejected during the 2011 outburst that gave rise to the radio emission not the ionization of a pre-existing circumbinary medium. They note that although the resolved pre-outburst H$\alpha$+[N II] emitting nebula surrounding T Pyx could be conceived to cause the rise ($t$$\sim$62-149 days) and fade of the radio light curve, the observed H$\alpha$ luminosities during the outburst are much too low to be consistent with this. In addition, they concluded that dense material in the immediate vicinity of the central binary should have similar characteristics to a stellar wind and, therefore, exhibit a partially optically thin radio spectrum while the observed spectrum during the rise appears to be completely optically thick \citep{nel12}.



\section{Conclusions}
We investigated the optical light curve of T Pyx in its 2011 outburst through compiling a database of SMEI and AAVSO observations.

 \begin{itemize}
    \item The SMEI light curve, providing unprecedented detail with high cadence data, was divided into four phases based on the idealised nova optical light curve: the initial rise; the pre-maximum halt; the final rise, and the early decline.

    \item A period of 1.44$\pm$0.05 days was the most strongly detected and was found in the interval from the first observation to the end of the pre-maximum halt phase, before the visual maximum. We compared this result to oscillations found in CVs and ascribed to accretion disc precession \citep{hir90}. Our observed $P_{orb}/P_{precession}$ is then 5.3$\%$ corresponding to $q$=0.125-0.15 which compares to $q$=0.2$\pm$0.03 derived by \citet{uth10}. Although the period is in line with that expected from studies of disc precession in CVs, we question however whether the disc would be present so early in the outburst. Such oscillations are present in some of the light curves derived by \citet{hil13} from TNR models and may be related to restructuring and rebalancing of the ejected envelope as it expands. We find no spectral variations related to the light curve periodicity however.
    \item The pre-maximum halt and subsequent dip in the SMEI light curve at $t$$\sim$22-24 days again resemble features in the light curves produced by \citeauthor{hil13}, with the latter possibly mirroring the shorter duration feature seen in other novae observed by SMEI \citep{hou10}. We note that this is coincident with a sharp transitory decline in H$\alpha$ flux and an equally sharp increase in that from Fe II.
    \end{itemize}

The spectra from the LT and SMARTS were investigated through each of the 4 phases of the optical light curve, in order to study the spectral evolution and investigate the physical causes of the variations of the light curve. We conclude, taking each phase in turn:

\begin{itemize}
\item {\bf Initial Rise ($t$=0.8-3.3 days).} The spectra show lines of high ionization species consistent with the presence of a high effective temperature pseudo-photosphere. The emission comes almost entirely from the continuum in this the ``fireball'' stage. The marked drop in the derived expansion velocity (4000 km s$^{-1}$ at $t$=0.8 days to $\sim$2000 km s$^{-1}$ at $t$=2.7 days) is consistent with the initial ejection in the form of a Hubble flow, but also resembles that noted in Type Ia SNe and ascribed to interaction with pre-outburst material.

\item {\bf Pre-Maximum Halt ($t$=3.6-13.7 days).} The subsequent change in behaviour of the derived $V_{ej}$ during the pre-maximum halt phase may suggest 2 different stages of mass loss: a short-lived phase occurring immediately after outburst and then followed by a more steadily evolving and higher mass loss phase. The fireball spectrum is maintained until $t$=8.6 days which is also the beginning of the typical iron curtain stage. Overall, the ionization/excitation and effective temperature of the underlying pseudo-photosphere appear to be decreasing through this phase, in line with basic models.

\item {\bf Final Rise ($t$=14.7-27.9 days).} The typical principal spectrum of CNe seems to be apparent and displays the O I flash beginning at $t$$\sim$17 days with the characteristics of the iron curtain phase still persisting. The gradual increase in $V_{ej}$, starting to appear in the final rise ($t$=14.8 days) and later, may be related to increasing ejection velocities from the central system, as proposed in other novae. The visual maximum at $t$=27.9 days seems to exhibit the lowest ionization lines, as expected.

\item {\bf Early Decline ($t$=27.9-90 days).} The Balmer lines began to have a double-peaked structure from $t$=42.7 days and this is shortly followed by the emergence of forbidden lines. The typical Orion spectrum is suggested to start at $t$$\sim$70 days. The emission line of N V at 4603\AA$ $ begins to emerge at $t$=73.7 days. The strong enhancement of N lines is associated with `nitrogen flaring' in the typical Orion spectrum stage.

\item {\bf Transition to the Nebular Phase ($t$=90-280 days).} By this time, the [O III] 5007\AA$ $ nebular and [Fe X] 6375\AA$ $ coronal lines have developed, the latter having been marginally detected at $t$=80 days while the rise in the X-ray light curve was detected at $t$=111 days. The last spectroscopic observation reported here at $t$=249.9 days in the nebular stage still exhibited [NII], He I, [Fe VII], and [Fe X] lines.
\end{itemize}

The overall spectral development of T Pyx is similar to that of CNe whose ejected mass is higher and velocity of ejection is lower than in typical RNe such as U Sco and RS Oph. We also found that in general the detected ionized elements are in line with those expected from the simple pseudo-photosphere models of \citet{bat89} at the same $\Delta$$B$ as shown in Table \ref{bat89} and Figure \ref{radio-x-lc}. An exception to this occurred near the end of the early decline phase where for e.g. N III and N V emission lines emission lines emerged earlier than would have been predicted.

In terms of the relationship of the optical development described here to that at other wavelengths, we found that:
\begin{itemize}
    \item The rise of the high frequency (33 GHz) radio emission was detected at $t$=7-15 days during the pre-maximum halt phase in the SMEI light curve, while X-rays were not yet detected.
    \item At $t$$\sim$45 days, where the [O III] 5007\AA$ $ was first present, the radio emission rose steeply while the X-rays were still undetectable.
    \item The rise in the X-ray emission, which began at $t$=111 days and rose to peak at $t$$\sim$144 days \citep{kuu11}, is consistent with the appearance of lines from highly ionized species such as [Ne III] 3869\AA$ $, [C III] 4364\AA$ $, N III 4640\AA$ $, found in the first blue spectrum ($t$=155.1 days) right after the seasonal gap.
    \item If the onset of the X-ray phase and the start of the final decline in the optical are related to the cessation of significant mass loss, this occurred at $t$$\sim$90-110 days.
    \item During the rise in X-rays, the radio flux kept increasing and peaked at around $t$$\sim$155 days for the highest frequency (37 GHz) corresponding to the middle of the plateau in the X-ray light curve where we clearly detected the coronal lines [Fe X] 6375\AA$ $ and [Fe VII] 6087\AA$ $ in our spectra ($t$=161.1 days).
    \item Although the appearance of the X-ray emission is in line with predictions of the emergence of the SSS from the simple \citet{bat89} model, we note that X-ray emission may be a mix of SSS and shocked circumstellar gas.
\end{itemize}

\acknowledgements We thank Kim Page for supplying the XRT light curve which was created from observations obtained
by the Swift nova-CV group (http://www.swift.-ac.uk/nova-cv/); Laura Chomiuk for supplying the EVLA radio data, and Bernard Jackson, Paul Hick, Andrew Buffington, and John M. Clover of the SMEI group at University California San Diego (UCSD) for assistance with work on the SMEI database. We also thank Sumner Starrfield and an anonymous referee for valuable comments on an earlier version of the manuscript.




\clearpage

\end{document}